\documentclass[review]{elsarticle}

\usepackage[table]{xcolor}  
\usepackage{mathtools}
\usepackage{amsmath}
\usepackage{amsmath,amssymb,amsthm}
\usepackage{graphicx}
\usepackage{subfig}
\usepackage[framemethod=tikz]{mdframed}
\usepackage{soul}

\usepackage{multirow}
\usepackage{commath}
\usepackage{url}

\usepackage{caption}
\usepackage{esvect}
\usepackage{booktabs} 
\usepackage{gensymb}
\usepackage{algcompatible}

\usepackage{algorithm}
\usepackage[noend]{algpseudocode}
\usepackage{setspace}

\makeatletter
\renewcommand{\ALG@beginalgorithmic}{\small}
\makeatother

\usepackage{tablefootnote}
\usepackage[margin=3cm]{geometry}

\makeatletter
\def\therule{\makebox[\algorithmicindent][l]{\hspace*{.5em}\vrule height .75\baselineskip depth .25\baselineskip}}%

\newtoks\therules
\therules={}
\def\appendto#1#2{\expandafter#1\expandafter{\the#1#2}}
\def\gobblefirst#1{
	#1\expandafter\expandafter\expandafter{\expandafter\@gobble\the#1}}%
\def\LState{\State\unskip\the\therules}
\def\LStatex{\Statex\unskip\the\therules}
\def\pushindent{\appendto\therules\therule}%
\def\popindent{\gobblefirst\therules}%
\def\printindent{\unskip\the\therules}%
\def\printandpush{\printindent\pushindent}%
\def\popandprint{\popindent\printindent}%

\algdef{SE}[WHILE]{While}{EndWhile}[1]
  {\printandpush\algorithmicwhile\ #1\ \algorithmicdo}
  {\popandprint\algorithmicend\ \algorithmicwhile}%
\algdef{SE}[FOR]{For}{EndFor}[1]
  {\printandpush\algorithmicfor\ #1\ \algorithmicdo}
  {\popandprint\algorithmicend\ \algorithmicfor}%
\algdef{S}[FOR]{ForAll}[1]
  {\printindent\algorithmicforall\ #1\ \algorithmicdo}%
\algdef{SE}[LOOP]{Loop}{EndLoop}
  {\printandpush\algorithmicloop}
  {\popandprint\algorithmicend\ \algorithmicloop}%
\algdef{SE}[REPEAT]{Repeat}{Until}
  {\printandpush\algorithmicrepeat}[1]
  {\popandprint\algorithmicuntil\ #1}%
\algdef{SE}[foreach]{Foreach}{EndForEach}[1]
  {\printandpush\algorithmicfor\ #1\ \algorithmicdo}
  {\popandprint\algorithmicend\ \algorithmicfor}%
\algdef{SE}[REPEAT]{Repeat}{Until}
  {\printandpush\algorithmicrepeat}[1]
  {\popandprint\algorithmicuntil\ #1}%
\algdef{SE}[IF]{If}{EndIf}[1]
  {\printandpush\algorithmicif\ #1\ \algorithmicthen}
  {\popandprint\algorithmicend\ \algorithmicif}%
\algdef{C}[IF]{IF}{ElsIf}[1]
  {\popandprint\pushindent\algorithmicelse\ \algorithmicif\ #1\ \algorithmicthen}%
\algdef{Ce}[ELSE]{IF}{Else}{EndIf}
  {\popandprint\pushindent\algorithmicelse}%
\algdef{SE}[PROCEDURE]{Procedure}{EndProcedure}[2]
   {\printandpush\algorithmicprocedure\ \textproc{#1}\ifthenelse{\equal{#2}{}}{}{(#2)}}%
   {\popandprint\algorithmicend\ \algorithmicprocedure}%
\algdef{SE}[FUNCTION]{Function}{EndFunction}[2]
   {\printandpush\algorithmicfunction\ \textproc{#1}\ifthenelse{\equal{#2}{}}{}{(#2)}}%
   {\popandprint\algorithmicend\ \algorithmicfunction}%
\makeatother

\usepackage{subfig}

\usepackage{footnote}

\newtheorem{mydef}{Definition}

\makeatletter
\def\BState{\State\hskip-\ALG@thistlm}
\makeatother

\usepackage{setspace}

\usepackage{breakcites}
\usepackage{enumerate}
\usepackage{makecell}

\usepackage{float}
\usepackage{listings}

\usepackage{tikz}
\usetikzlibrary{trees}
\usepackage{paralist}
\usepackage{ragged2e}
\usepackage[framemethod=tikz]{mdframed}

\usetikzlibrary{positioning,arrows.meta}

\definecolor{arrowblue}{RGB}{98,145,224}

\usetikzlibrary{arrows.meta}
\tikzset{%
	>={Latex[width=2mm,length=2mm]},
	base/.style = {rectangle, rounded corners, 
		minimum width=11cm, minimum height=1cm,
		text centered, font=\sffamily},
	activityStarts/.style = {base, fill=blue!30},
	startstop/.style = {base, fill=red!30},
	activityRuns/.style = {base, fill=green!30},
	activityRuns2/.style = {base, fill=magenta!30},
	activityRuns3/.style = {base, fill=yellow!30},
	activityRuns4/.style = {base, fill=violet!30},
	activityRuns5/.style = {base, fill=orange!30},
	activityRuns6/.style = {base, fill=cyan!30},
	activityRuns7/.style = {base, fill=purple!30},
	activityRuns8/.style = {base, fill=gray!30},
	activityRuns9/.style = {base, fill=orange!30},
	activityRuns10/.style = {base, fill=blue!30},
	process/.style = {base, minimum width=10 cm, fill=orange!15,
		font=\ttfamily},
}
\usepackage{verbatim}
\usepackage{smartdiagram}
\usepackage{metalogo}

\usepackage{footnote}

\makeatletter
\def\BState{\State\hskip-\ALG@thistlm}
\makeatother

\usepackage{setspace}

\usepackage{breakcites}
\usepackage{enumerate}
\usepackage{makecell}

\usepackage{float}
\usepackage{listings}

\usepackage{tikz}
\usetikzlibrary{trees}
\usepackage{paralist}
\usepackage{ragged2e}
\usepackage[framemethod=tikz]{mdframed}

\usetikzlibrary{positioning,arrows.meta}

\definecolor{arrowblue}{RGB}{98,145,224}

\usepackage{lineno,hyperref}
\usepackage{cleveref}
\usepackage{setspace}
\modulolinenumbers[5]

\journal{Journal of Computers $\&$ Security}

\begin{document}

\begin{frontmatter}

\title{
An Efficient and Scalable Privacy Preserving Algorithm for Big Data and Data Streams}

\author[mymainaddress,mysecondaryaddress]{M.A.P.~Chamikara
	\corref{mycorrespondingauthor}}
\cortext[mycorrespondingauthor]{Corresponding author}
\ead{pathumchamikara.mahawagaarachchige@rmit.edu.au}

\author[mymainaddress]{P.~Bertok}
\author[mysecondaryaddress]{D.~Liu}
\author[mysecondaryaddress]{S.~Camtepe}
\author[mymainaddress]{I.~Khalil}

\address[mymainaddress]{School of Science, RMIT University, Australia}
\address[mysecondaryaddress]{CSIRO Data61, Australia}
 
\begin{abstract}
\begin{mdframed}[backgroundcolor=green!50,rightline=false,leftline=false]
\centering 
The published article can be found at \url{https://doi.org/10.1016/j.cose.2019.101570}
\end{mdframed}

A vast amount of valuable data is produced and is becoming available for analysis as a result of advancements in smart cyber-physical systems. The data comes from various sources, such as healthcare, smart homes, smart vehicles, and often includes private, potentially sensitive information that needs appropriate sanitization before being released for analysis. The incremental and fast nature of data generation in these systems necessitates scalable privacy-preserving mechanisms with high privacy and utility.  However, privacy preservation often comes at the expense of data utility. We propose a new data perturbation algorithm, SEAL (Secure and Efficient data perturbation Algorithm utilizing Local differential privacy), based on Chebyshev interpolation and Laplacian noise, which provides a good balance between privacy and utility with high efficiency and scalability. Empirical comparisons with existing privacy-preserving algorithms show that SEAL excels in execution speed, scalability, accuracy, and attack resistance. SEAL provides flexibility in choosing the best possible privacy parameters, such as the amount of added noise, which can be tailored to the domain and dataset.

\end{abstract}


\end{frontmatter}


\section{Introduction}
Smart cyber-physical systems (SCPS) such as smart vehicles, smart grid, smart healthcare systems, and smart homes are becoming widely popular due to massive technological advancements in the past few years. These systems often interact with the environment to collect data mainly for analysis, e.g.
to allow life activities  to be more intelligent, efficient, and reliable~\cite{de2016iot}. Such data often includes sensitive details, but sharing confidential information with third parties can lead to a privacy breach. From our perspective, privacy can be considered as ``Controlled Information Release'' \cite{bertino2008survey}. We can define a privacy breach as the release of private/confidential information to an untrusted environment. However, sharing the data with external parties may be necessary for data analysis, such as data mining and machine learning.  Smart cyber-physical systems must have the ability to share information while limiting the disclosure of private information to third parties.   Privacy-preserving data sharing and privacy-preserving data mining face significant challenges because of the size of the data and the speed at which data are produced.  Robust, scalable, and efficient solutions are needed to preserve the privacy of big data and data streams generated by SCPS ~\cite{zhang2016privacy,wen2018scheduling}.  Various solutions for privacy-preserving data mining (PPDM) have been proposed for data sanitization; they aim to ensure confidentiality and privacy of data during data mining ~\cite{xue2011distributed, backes2016profile, yang2017efficient, vatsalan2017privacy}. 

The two main approaches of PPDM are data perturbation  ~\cite{chen2005random, chen2011geometric} and encryption ~\cite{li2015towards,kerschbaum2017searchable}.  Although encryption provides a strong notion of security, due to its high computation complexity~\cite{gai2016privacy} it can be impractical for PPDM of SCPS-generated big data and data streams. Data perturbation, on the other hand, applies certain modifications such as randomization and noise addition to the original data to preserve privacy~\cite{agrawal2005framework}. These modification techniques are less complex than cryptographic mechanisms~\cite{xu2014building}. Data perturbation mechanisms such as noise addition~\cite{muralidhar1999general} and randomization~\cite{fox2015randomized} provide efficient solutions towards PPDM.  However, the utility of
perturbed data cannot be 100\% as data perturbation applies modifications to the original data, and the ability to infer knowledge from the perturbed data can result in a certain level of privacy leak as well. A privacy model~\cite{machanavajjhala2015designing} describes the limitations to the utility and privacy of a perturbation mechanism. Examples of such earlier privacy models include $k-anonymity$~\cite{niu2014achieving,zhang2016designing} and $l-diversity$~\cite{machanavajjhala2006diversity}. However, it has been shown that older privacy models are defenseless against certain types of attacks, such as minimality attacks~\cite{zhang2007information}, composition attacks~\cite{ganta2008composition} and foreground knowledge~\cite{wong2011can} attacks. Differential privacy (DP)  is a privacy model that provides a robust solution to these issues by rendering maximum privacy via minimizing the chance of private data leak~\cite{dwork2009differential, mohammed2011differentially,friedman2010data,wang2015outsourcing}. Nevertheless, current DP mechanisms fail for small databases and have limitations on implementing efficient solutions for data streams and big data. When the database is small, the utility of DP mechanisms diminishes due to insufficient data being available for a reasonable estimation of statistics~\cite{qin2016heavy}. At the other end of the scale, when the database is very large or continuously growing like in data streams produced by SCPS, the information leak of DP mechanisms is high due to the availability of too much information~\cite{kellaris2014differentially}. Most perturbation mechanisms tend to leak information when the data is high-dimensional, which is a consequence of the dimensionality curse~\cite{aggarwal2005k}. Moreover, the significant amount of randomization produced by certain DP algorithms results in low data utility.  Existing perturbation mechanisms often ignore the connection between utility and privacy, even though improvement of one leads to deterioration of the other~\cite{mivule2013comparative}. Furthermore, the inability to efficiently process high volumes of data and data streams makes the existing methods unsuitable for privacy-preservation in smart cyber-physical systems. New approaches which can appropriately answer the complexities in privacy preservation of SCPS generated data are needed. 

The main contribution of this paper is a robust and efficient privacy-preserving algorithm for smart cyber-physical systems, which addresses the issues existing perturbation algorithms have. Our solution, SEAL (Secure and Efficient data perturbation Algorithm utilizing Local differential privacy), employs polynomial interpolation and notions of differential privacy.  SEAL is a linear perturbation system based on Chebyshev polynomial interpolation, which allows it to work faster than comparable methods. We used generic datasets retrieved from the UCI data repository\footnote{https://archive.ics.uci.edu/ml/index.php} to evaluate SEAL's efficiency, scalability, accuracy, and attack resistance. The results indicate that SEAL performs well at privacy-preserving data classification of big data and data streams. SEAL outperforms existing alternative algorithms in efficiency, accuracy, and data privacy, which makes it an excellent solution for smart system data privacy preservation.

The rest of the paper is organized as follows. Section \ref{relwork} provides a summary of existing related work. The fundamentals of the proposed method are briefly discussed in Section \ref{fndmntls}. Section \ref{methodology} describes the technical details of SEAL. Section \ref{expresults} presents the experimental settings and provides a comparative analysis of the performance and security of PABIDOT.  The results are discussed in Section \ref{discussion}, and the paper is concluded in Section \ref{conclusion}. Detailed descriptions of the underlying concepts of SEAL are given in the Appendices.

\section{Related Work}
\label{relwork}
Smart cyber-physical systems (SCPS) have become an important part of the IT landscape. Often these systems include IoT devices that allow effective and easy acquisition of data in areas such as healthcare, smart cities, smart vehicles, and smart homes~\cite{de2016iot}. Data mining and analysis are among the primary goals of collecting data from SCPS. The infrastructural extensions of SCPSs have contributed to the exponential growth in the number of IoT sensors, but security is often overlooked, and the devices become a source of privacy leak. The security and privacy concerns of big data and data streams are not entirely new, but require constant attention due to technological advancements of the environments and the devices used~\cite{kieseberg2018security}. Confidentiality, authentication, and authorization are just a few of the concerns~\cite{balandina2015iot, sridhar2012cyber,fernando2016consumer}.  Many studies have raised the importance of privacy and security of SCPS due to their heavy use of personally identifiable information (PII)~\cite{liu2012cyber}.  Controlling access via authentication~\cite{bertino2016data}, attribute-based encryption~\cite{wang2014performance}, temporal and location-based access control~\cite{bertino2016data} and employing constraint-based protocols~\cite{kirkham2015privacy} are some examples of improving privacy of SCPS. 

In this paper, our primary target is maintaining privacy when sharing and mining data produced by SCPSs, and the focus is on ``controlled information release''. Literature shows different attempts to impose constraints on data release and analysis in order to preserve privacy~\cite{aldeen2015comprehensive}. Data encryption and data perturbation-based solutions have proven to be more viable for privacy-preserving data publishing and analysis than methods based on authentication and authorization~\cite{verykios2004state}. Some recent examples for encryption based privacy-preserving approaches for cloud computing include PPM ~\cite{razaque2017privacy}, Sca-PBDA~\cite{wu2016scalable} and TDPP~\cite{razaque2016triangular}, which provide scalable privacy-preserving data processing infrastructures. However, cryptographic mechanisms are less popular in PPDM for ``controlled information release'' due to the high computational complexity, hence not suitable for resource-constrained devices.  Perturbing the instances of the original data by introducing noise or randomization is called input perturbation~\cite{agrawal2000privacy, aldeen2015comprehensive}, whereas perturbing the outputs of a query or analysis using noise addition or rule hiding is called output perturbation. Unidimensional perturbation and multidimensional perturbation  \cite{agrawal2000privacy,datta2004random,liu2006random,zhong2012mu} are the two main types of input perturbation classes. Examples for unidimensional perturbation include but are not limited to additive perturbation \cite{muralidhar1999general}, randomized response \cite{du2003using}, and swapping \cite{estivill1999data}. Condensation \cite{aggarwal2004condensation}, random rotation \cite{chen2005random}, geometric perturbation \cite{chen2011geometric}, random projection \cite{liu2006random}, and hybrid perturbation are types of multidimensional perturbation~\cite{aldeen2015comprehensive}. Microaggregation\cite{domingo2002practical} can be considered as a hybrid perturbation mechanism that possesses both unidimensional and multidimensional perturbation capabilities.

As privacy models evolved, the limits of privacy imposed by particular mechanisms were evaluated~\cite{machanavajjhala2015designing}, and new models were defined to overcome the issues of their predecessors. For example, $l-diversity$~\cite{machanavajjhala2006diversity} was defined to overcome the shortcomings of  $k-anonymity$~\cite{niu2014achieving}, then $(\alpha, k)-anonymity$~\cite{wong2006alpha}, $t-closeness$~\cite{li2007t} were proposed as further improvements. However, all these models eventually exhibited vulnerabilities to different attacks such as minimality~\cite{zhang2007information}, composition~\cite{ganta2008composition} and foreground knowledge~\cite{wong2011can} attacks. Moreover, these models were not scalable enough to address the curse of dimensionality presented by big data and data streams~\cite{aggarwal2005k,cao2011castle}, hence resulted in higher privacy leak~\cite{aggarwal2005k}. In recent years differential privacy (DP) has drawn much attention as a powerful privacy model due to its inherent strong privacy guarantee. Differential privacy that is achieved via output perturbation is known as global differential privacy (GDP), whereas differential privacy achieved using input perturbation is known as local differential privacy (LDP). Laplacian mechanism and randomized response are two of the most frequently employed data perturbation methods used to achieve GDP and LDP~\cite{dwork2014algorithmic,fox2015randomized}. LDP permits full or partial data release allowing the analysis of the perturbed data~\cite{dwork2008differential,kairouz2014extremal}, while GDP requires a trusted curator who enforces differential privacy by applying noise or randomization on results generated by running queries or by analysis of the original data~\cite{dwork2008differential}. Nevertheless, LDP algorithms are still at a rudimentary stage when it comes to full or partial data release of real-valued numerical data, and the complexity of selecting the domain of randomization with respect to a single data instance is still a challenge~\cite{erlingsson2014rappor,cormode2018privacy,qin2016heavy}. Consequently, existing DP mechanisms are not suitable for differentially private data release.

Two important characteristics that determine the robustness of a particular perturbation mechanism are the ability to (1) protect against data reconstruction attacks and (2) perform well when high dimensional datasets and data streams are introduced. A data reconstruction attack tries to re-identify the individual owners of the data by reconstructing the original dataset from the perturbed dataset. Data reconstruction attacks are normally custom built for different perturbation methods using the properties of the perturbation mechanisms to restore the original data systematically.  Different perturbation scenarios are vulnerable to different data reconstruction attacks. Principal component analysis \cite{wold1987principal}, maximum likelihood estimation \cite{scholz2006maximum}, known I/O attack \cite{aggarwal2008general}, ICA attack \cite{chen2005privacy} and known sample attack \cite{aggarwal2008general} are some examples of common data reconstruction attacks. For example, additive perturbation is vulnerable to principal component analysis \cite{huang2005deriving} and maximum likelihood estimation~\cite{huang2005deriving}, whereas multiplicative data perturbation methods are vulnerable to known input/output (I/O) attacks, known sample attacks and ICA attacks. These reconstruction attacks exploit the high information leak due to the dimensionality curse associated with high-dimensional data. In addition to providing extra information to the attacker, high-dimensional data also exponentially increases the amount of necessary computations~\cite{aggarwal2005k, machanavajjhala2006diversity, chen2005random,chen2011geometric}.

The literature shows methods that try to provide privacy-preserving solutions for data streams and big data by addressing the dimensionality curse in data streams. Recent attempts include a method proposing steered microaggregation to anonymize a data stream to achieve $k-anonymity$~\cite{domingo2017steered}, but the problem of information leak inherent in $k-anonymity$ in case of high dimensional data can be a shortcoming of the method. Xuyun Zhang et al.~\cite{zhang2015proximity} introduced a scalable data anonymization with MapReduce for big data privacy preservation. The proposed method first splits an original data set into partitions that contain similar data records in terms of quasi-identifiers and then locally recodes the data partitions by the proximity-aware agglomerative clustering algorithm in parallel which limits producing sufficient utility for data streams such as produced by SCPS. Further, the requirement of advanced processing capabilities precludes its application to resource-constrained devices. Data condensation is another contender for data stream privacy preservation~\cite{ aggarwal2008static}. The problem, in this case, is that when the method parameters are set to achieve high accuracy (using small spatial locality), the privacy of the data often suffers. $P^2RoCAl$ is a privacy preservation algorithm that provides high scalability for data streams and big data composed of millions of data records. $P^2RoCAl$ is based on data condensation and random geometric transformations. Although $P^2RoCAl$ provides linear complexity for the number of tuples, the computational complexity increases exponentially for the number of attributes~\cite{chamikaraprocal}. PABIDOT is a scalable privacy preservation algorithm for big data ~\cite{chamikara2019efficient}. PABIDOT comes with a  new privacy model named $\Phi-separation$, which facilitates full data release with optimum privacy protection. PABIDOT can process millions of records efficiently as it has linear time complexity for the number of instances. However, PABIDOT also shows exponential time complexity for the number of attributes of a dataset~\cite{chamikara2019efficient}.   Other examples include the use of a Naive Bayesian Classifier for private data streams~\cite{xu2008privacy}, and the method to efficiently and effectively track the correlation and autocorrelation structure of multivariate streams and leverage it to add noise to preserve privacy~\cite{li2007hiding}. The latter methods are also vulnerable to data reconstruction attacks such as principal component analysis-based attacks.

The complex dynamics exhibited by SCPS require efficient privacy preservation methods scalable enough to handle exponentially growing databases and data streams such as IoT stream data. Existing privacy-preserving mechanisms have difficulties in maintaining the balance between privacy and utility while providing sufficient efficiency and scalability. This paper tries to fill the gap, and proposes a privacy-preserving algorithm for SCPS which solves the existing issues.

\section{Fundamentals}
\label{fndmntls}
In this section, we provide some background and discuss the fundamentals used in the proposed method (SEAL). Our approach is generating a privacy-preserved version of the dataset in question, and allowing only the generated dataset to be used in any application. We use Chebyshev interpolation based on least square fitting to model a particular input data series, and the model formation is subjected to noise addition using the Laplacian mechanism used in differential privacy. The noise integrated model is then used to synthesize a perturbed data series which approximate the properties of the original input data series. 

\subsection{Chebyshev Polynomials of the First Kind}
For the interpolation of the input dataset, we use Chebyshev Polynomials of the First Kind. These polynomials are a set of orthogonal polynomials as given by Definition \ref{defp1} (available in \ref{chebyshevapp}) ~\cite{mason2002chebyshev}. Chebyshev polynomials are a sequence of orthonormal polynomials that can be defined recursively. Polynomial approximation and numerical integration are two of the areas where Chebyshev polynomials are heavily used~\cite{mason2002chebyshev}.   More details on Chebyshev polynomials of the first kind can be found in \ref{chebyshevapp}.

\subsection{Least Squares Fitting}
Least squares fitting (LSF) is a mathematical procedure which minimizes the sum of squares of the offsets of the points from the curve to find the best-fitting curve to a given set of points. We can use vertical least squares fitting which proceeds by finding the sum of squares of the vertical derivations $R^2$ (refer Equation \ref{vlsf} in \ref{lsfapp}) of a set of $n$ data points~\cite{weisstein2002least}. To generate a linear fit considering  $f(x)=mx+b$, we can minimize the expression of squared error between the estimated values and the original values (refer Equation \ref{linfit}), which proceeds to obtaining the linear system shown in Equation \ref{matrixform22} (using Equations \ref{solpart1} and \ref{solpart2}). We can solve Equation \ref{matrixform22} to find values of $a$ and $b$ to obtain the corresponding linear fit of $f(x)=mx+b$ for a given data series. 

\begin{equation}
\begin{bmatrix}
    b \\
    m 
   \end{bmatrix}
   =
    \begin{bmatrix}
    n &  \left(\sum_{i=1}^n x_i \right) \\
   \left(\sum_{i=1}^n x_i \right) & \left(\sum_{i=1}^n x_i^2 \right) 
   \end{bmatrix}^{-1}
   \begin{bmatrix}
    \sum_{i=1}^n y_i \\
    \sum_{i=1}^n x_i y_i 
   \end{bmatrix}
    \label{matrixform22}
\end{equation}
\subsection{Differential Privacy}

Differential Privacy (DP) is a privacy model that defines the bounds to how much information can be revealed to a third party or adversary about someone's data being present or absent in a particular database. Conventionally, $\epsilon$ (epsilon) and $\delta$ (delta) are used to denote the level of privacy rendered by a randomized privacy preserving algorithm ($M$) over a particular database ($D$).   Let us take two $x$ and $y$ adjacent datasets of $D$, where $y$ differs from $x$ only by one person.  Then $M$ satisfies ($\epsilon$, $\delta$)-differential privacy if  Equation \eqref{dpeq} holds.

\emph{ Privacy Budget and Privacy Loss ($\epsilon$):} $\epsilon$ is called the privacy budget that provides an insight into the privacy loss of a DP algorithm. When the corresponding $\epsilon$ value of a particular differentially private algorithm $A$ is increased, the amount of noise or randomization applied by $A$ on the input data is decreased.   The higher the value of $\epsilon$, the higher the privacy loss. 

\emph{Probability to Fail a.k.a. Probability of Error ($\delta$):} $\delta$ is the parameter that accounts for "bad events" that might result in high privacy loss; $\delta$ is the probability of the output revealing the identity of a particular individual, which can happen $n \times \delta$ times where $n$ is the number of records. To minimize the risk of privacy loss, $n \times \delta$ has to be maintained at a low value. For example, the probability of a bad event is 1\% when $\delta=\frac{1}{100\times n}$.

\begin{mydef}
A randomized algorithm $M$ with domain $N^{|X|}$ and
range $R$ is ($\epsilon$, $\delta$)-differentially private for $\delta \geq 0$,  if for every adjacent $x$, $y$ $\in$ $N^{|X|}$
 and for any subset $S \subseteq R$
\end{mydef}

\begin{equation}
Pr[(M(x) \in S)] \leq \exp(\epsilon) Pr[(M(y) \in S)] + \delta
\label{dpeq}
\end{equation}







\subsection{Global vs. Local Differential Privacy}

Global differential privacy (GDP) and local differential privacy (LDP) are the two main approaches to differential privacy.  In the GDP setting, there is a trusted curator who applies carefully calibrated random noise to the real values returned for a particular query. The GDP setting is also called the trusted curator model~\cite{chan2012differentially}. Laplace mechanism and Gaussian mechanism~\cite{dwork2014algorithmic} are two of the most frequently used noise generation methods in GDP~\cite{dwork2014algorithmic}.  A randomized algorithm, $M$ provides $\epsilon$-global differential privacy if for any two adjacent datasets $x,y$ and $S \subseteq R$, $Pr[(M(x) \in S)] \leq \exp(\epsilon) Pr[(M(y) \in S)] + \delta$ (i.e.  Equation \eqref{dpeq} holds). On the other hand, LDP eliminates the need of a trusted curator by randomizing the data before the curator can access them. Hence, LDP  is also called the untrusted curator model ~\cite{kairouz2014extremal}.  LDP can also be used by a trusted party to randomize all records in a database at once. LDP algorithms may often produce too noisy data, as noise is applied commonly to achieve individual record privacy. LDP is considered to be a strong and rigorous notion of privacy that provides plausible deniability. Due to the above properties, LDP is deemed to be a state-of-the-art approach for privacy-preserving data collection and distribution. A randomized algorithm $A$ provides $\epsilon$-local differential privacy if  Equation \eqref{ldpeq} holds ~\cite{erlingsson2014rappor}.

\begin{mydef}
A randomized algorithm $A$  satisfies $\epsilon$-local differential privacy, if for all pairs of inputs $v_1$ and $v_2$, for all $Q \subseteq Range(A)$ and for ($\epsilon \geq 0$)  Equation \eqref{ldpeq} holds. $Range(A)$ is the set of all possible outputs of the randomized algorithm $A$.
\end{mydef}

\begin{equation}
Pr[A(v_1) \in Q] \leq \exp(\epsilon) Pr[A(v_2) \in Q]
\label{ldpeq}
\end{equation}

\subsection{Sensitivity}

Sensitivity is defined as the maximum influence that a single individual data item can have on the result of a numeric query. Consider a function $f$, the sensitivity  ($\Delta f$) of $f$ can be given as in Equation \eqref{seneq} where x and y are two neighboring databases (or in LDP, adjacent records) and $\lVert . \rVert_1$ represents the $L1$ norm of a vector~\cite{wang2016using}. 

\begin{equation}
\Delta f=max\{\lVert f(x)-f(y) \rVert_1\}
\label{seneq}
\end{equation}

\subsection{Laplace Mechanism}

The Laplace mechanism is considered to be one of the most generic mechanisms to achieve differential privacy~\cite{dwork2014algorithmic}. Laplace noise can be added to a function output ($F(D)$) as given in Equation \eqref{diffeq21} to produce a differentially private output. $\Delta f$ denotes the sensitivity of the function $f$. In the local differentially private setting, the scale of the Laplacian noise is equal to $\Delta f/\epsilon$, and the position is the current input value ($F(D)$).

\begin{equation}
    PF(D)= F(D)+Lap(\frac{\Delta f}{\epsilon})
\label{diffeq}
\end{equation}

\begin{equation}
    PF(D)= \frac{\epsilon}{2\Delta f}e^{-\frac{|x-F(D)|}{\Delta F}}
\label{diffeq21}
\end{equation}

\section{Our Approach}
\label{methodology}
\begin{figure}[H]
	\centering
	\scalebox{0.8}{
		\includegraphics[width=1\textwidth, trim=0cm 0cm 0cm
		0cm]{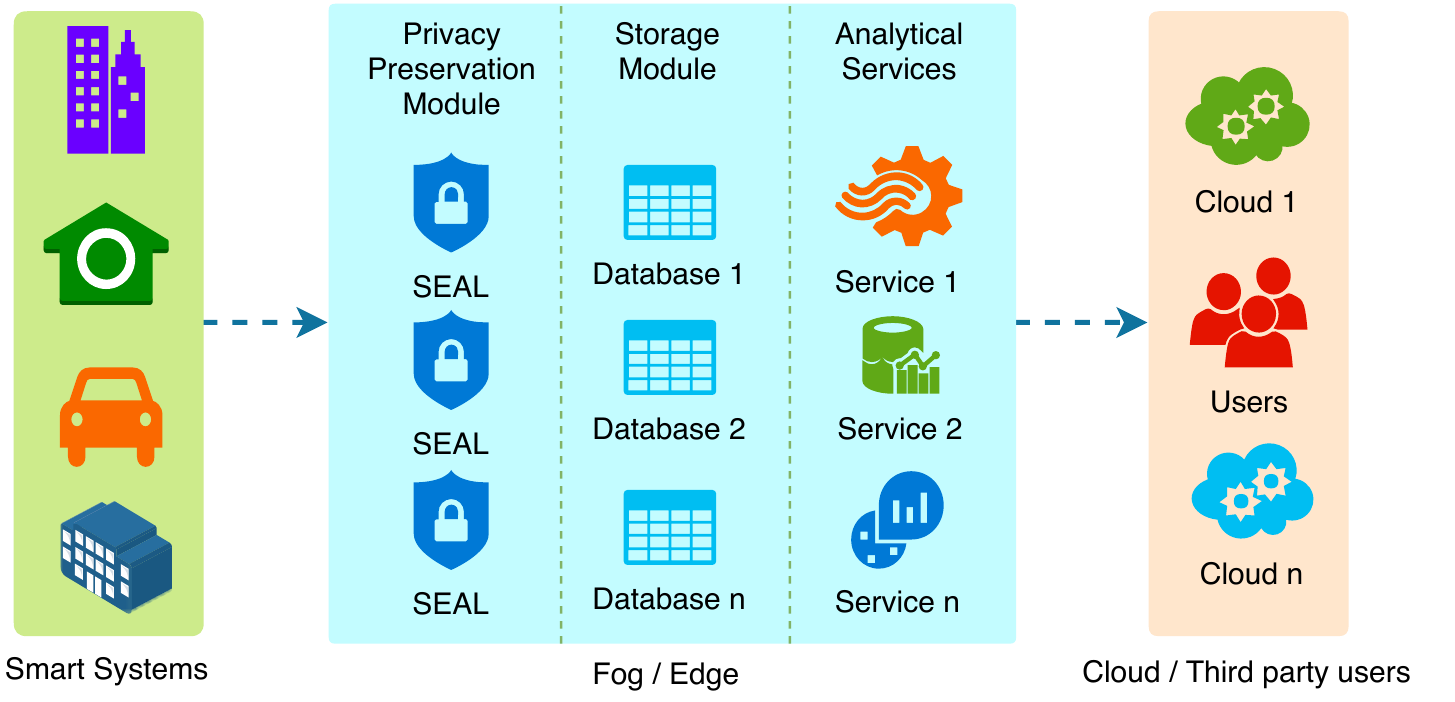}
	}
	\caption{Arrangement of SEAL in a smart system environment. In this setting, we assume that the original data are perturbed before reaching the storage devices.  Any public or private services will have access only to the perturbed data.}
	\label{sealtposition}
\end{figure}

The proposed method, SEAL, is designed to preserve the privacy of big data and data streams generated by systems such as smart cyber-physical systems. One of our aims was balancing privacy and utility, as they may adversely affect each other. For example, the spatial arrangement of a dataset can potentially contribute to its utility in data mining, as the results generated by the analysis mechanisms such as data classification and clustering are often influenced by the spatial arrangement of the input data. However, the spatial arrangement can be affected when privacy mechanisms apply methods like randomization. In other words, while data perturbation mechanisms improve privacy, at the same time they may reduce utility. Conversely, an increasing utility can detrimentally affect privacy. To address these difficulties, SEAL processes the data in three steps: (1) determine the sensitivity of the dataset to calibrate how much random noise is necessary to provide sufficient privacy, (2) conduct polynomial interpolation with calibrated noise to approximate a noisy function over the original data, and (3) use the approximated function to generate perturbed data.  These steps guarantee that SEAL applies enough randomization to preserve privacy while preserving the spatial arrangement of the original data.  SEAL uses polynomial interpolation accompanied by noise addition, which is calibrated according to the instructions of differential privacy. We use the first four orders of the Chebyshev polynomial of the first kind in the polynomial interpolation process. Then we calibrate random Laplacian noise to apply a stochastic error to the interpolation process, in order to generate the perturbed data. Figure \ref{sealtposition} shows the integration of SEAL in the general purpose data flow of SCPS. As shown in the figure, the data perturbed by the SEAL layer comes directly from the SCPS. That means that the data in the storage module has already gone through SEAL's privacy preservation process and does not contain any original data.

\begin{figure}[H]
	\centering
	\scalebox{1}{
		\includegraphics[width=1\textwidth, trim=0cm 0cm 0cm
		0cm]{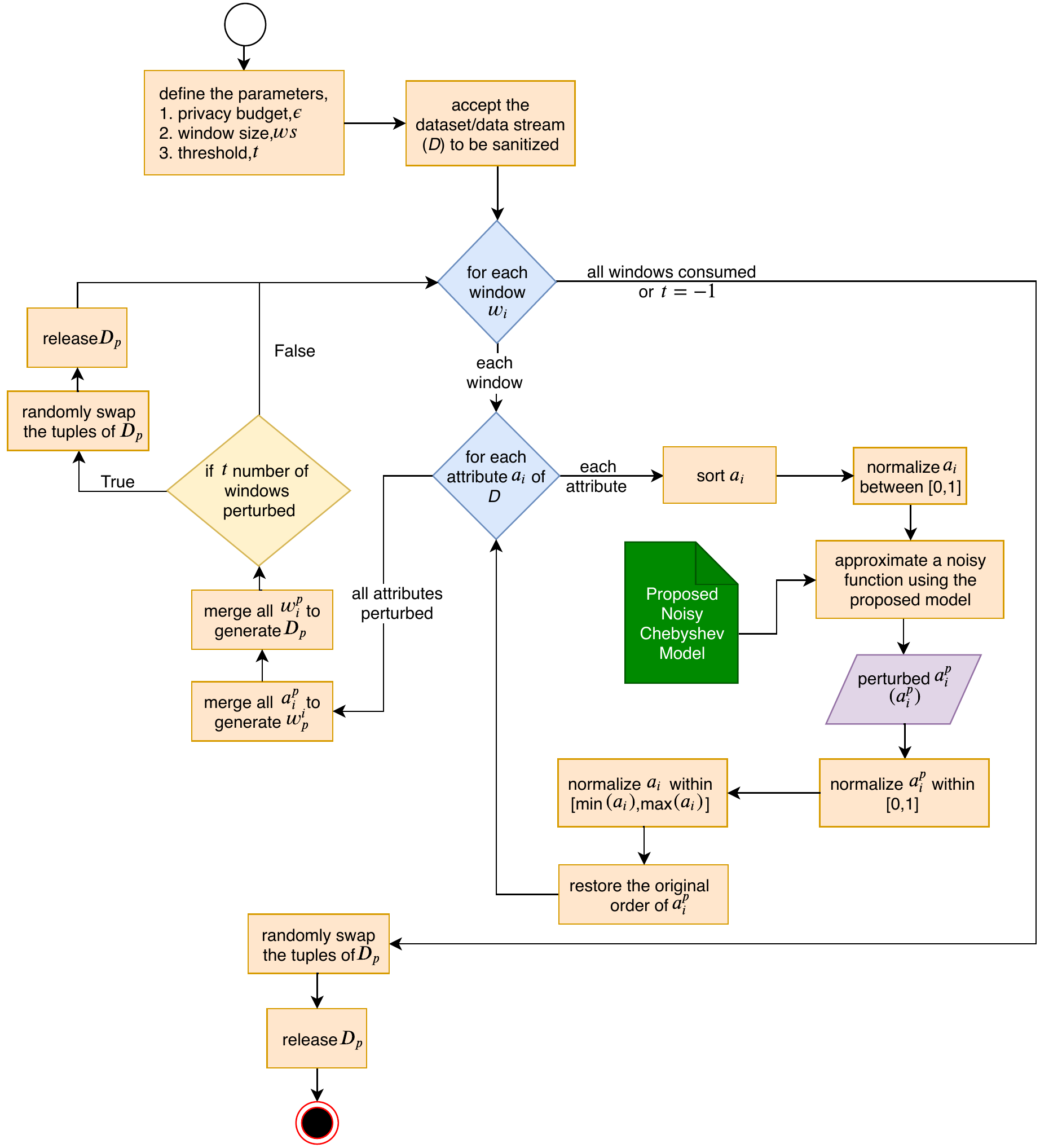}
	}
	\caption{The basic flow of SEAL. The users can calibrate the level of privacy using the privacy budget parameter ($\epsilon$). The smaller the $\epsilon$, the higher the privacy. It is recommended to use an $\epsilon$  in the interval  $(0,10)$, which is considered to be the default range to provide a sufficient level of privacy. }
	\label{flowchart}
\end{figure}

Figure \ref{flowchart} shows the flow of SEAL where the proposed noisy Chebyshev model (represented by a green note) is used to approximate each of the individual attributes of a particular input dataset or data stream. The approximated noisy function is used to synthesize perturbed data, which is then subjected to random tuple shuffling to reduce the vulnerability to data linkage attacks.

\subsection{Privacy-Preserving Polynomial Interpolation for Noisy Chebyshev Model Generation}

We approximate an input data series (experimental data) by a functional expression with added randomization in order to inherit the properties of differential privacy. For approximation, our method uses the first four orders of Chebyshev polynomials of the first kind. We systematically add calibrated random Laplacian noise in the interpolation process, i.e. apply randomization to the approximation. Then we use the approximated function to re-generate the dataset in a privacy-preserving manner. We can denote an approximated function $\hat{f}$ of degree $(m-1)$ using Equation \ref{approxfunc11}, where the degree of $(\varphi_k)$ is $k-1$. For the approximation, we consider the root mean square error (RMSE) $E$ between the estimated values and the original values (refer to Equation \ref{rmse}). We use the first four Chebyshev polynomials of the first kind for the approximation, which limits the number of coefficients to four (we name the coefficients as $a_1, a_2, a_3,$ and $a_4$). Now we can minimize $E$ (the RMSE) to obtain an estimated function $\hat{f}^*(x)$, thus seeking to minimize the squared error $M(a_1,a_2,a_3,a_4)$. For more details refer to Equation \ref{minrmse}, where $a_1,a_2,\dots,a_m$ are coefficients  and $\varphi_1(x), \varphi_2(x),\dots, \varphi_m(x)$ are Chebyshev polynomials of first kind. 

\begin{equation}
 \hat{f}(x)=a_1\varphi_1(x)+a_2\varphi_2(x)+\dots+a_m\varphi_m(x)   
 \label{approxfunc11}
\end{equation}

\subsubsection{Introducing privacy to the approximation process utilizing differential privacy (the determination of the sensitivity and the position of Laplacian noise)}

We apply the notion of differential privacy to the private data generation process by introducing randomized Laplacian noise to the root mean square error (RMSE) minimization process. Random Laplacian noise introduces a calibrated randomized error in deriving the values for $a_1,a_2,a_3,$ and $a_4$ with an error (refer to Equations \ref{rapartial1},  \ref{rapartial2},  \ref{rapartia3} and  \ref{rapartia4}). We add Laplacian noise with a sensitivity of 1, as the input dataset is normalized within the bounds of 0 and 1, which restricts the minimum output to 0 and maximum output to 1 (refer to Equation \eqref{sensitivity}). We select the position of Laplacian noise to be 0, as the goal is to keep the local minima of RMSE around 0. We can factorize the noise introduced squared error minimization equations to form a linear system which can be denoted by Equation \ref{linsys1}. $C$ is the coefficient matrix obtained from the factorized expressions, $A$ is the coefficient vector obtained from $M$, and  $B$ is the constant vector obtained from the factorized expressions (refer Equations \ref{rapartial13}, \ref{rapartial23}, \ref{rapartial33}, and \ref{rapartial43} where $m_{ij}$ denote the coefficients and $b_i$ denote the constants). 

\begin{equation}
\resizebox{0.08\hsize}{!}{$
    CA=B
    $}
    \label{linsys1}
\end{equation}

Where,
\begin{equation}
\resizebox{0.35\hsize}{!}{$
    C=
    \begin{bmatrix}
    m_{11} &  m_{12} & m_{13} & m_{14} \\
    m_{21} & m_{22} & m_{23} & m_{24} \\
    m_{31} & m_{32} & m_{33} & m_{34} \\
    m_{41} & m_{42} & m_{43} & m_{44}
    
   \end{bmatrix}
   $}
    \label{matM1}
\end{equation}

\begin{equation}
A=\left[a_1, a_2, a_3, a_4\right]^T
    \label{vectorA1}
\end{equation}

\begin{equation}
B=\left[b_1, b_2, b_3, b_4\right]^T
\label{vectorB1}
\end{equation}

Now we solve the corresponding linear system (formed using  Equations \ref{matM}-\ref{vectorB}), to obtain noisy values for $a_1, a_2, a_3$, and $a_4$ in order to approximate the input data series with a noisy function. The results will differ each time we calculate the values for  $a_1, a_2, a_3$, and $a_4$ as we have randomized the process of interpolation by adding randomized Laplacian noise calibrated using a user-defined $\epsilon$ value. 

\subsection{Algorithmic Steps of SEAL for Static Data and Data Streams}

Algorithm \ref{privatealgo} presents the systematic flow of steps in randomizing the data to produce a privacy-preserving output. The algorithm accepts input dataset ($D$), privacy budget $\epsilon$ (defined in Equation \ref{diffeq2}), window size ($ws$) and threshold ($t$) as the input parameters. The window size defines the number of data instances to be perturbed in one cycle of randomization. The window size of a data stream is essential to maintain the speed of the post-processing analysis/modification (e.g. data perturbation, classification, and clustering) done to the data stream~\cite{hammad2003scheduling}. For static data sets, the threshold is maintained with a default value of $-1$. For a static dataset,  $t=-1$ ignores that a specific number of perturbed windows need to be released before the whole dataset is completed.   In the case of data streams, the window size ($ws$) and the threshold $t$ are useful as $ws$ can be maintained as a data buffer and $t$ can be specified with a certain number to let the algorithm know that it has to release every $t$ number of processed windows. Maintaining $t$ is important for data streams because data streams are growing infinitely in most cases, and the algorithm makes sure that the data is released in predefined intervals. 

According to conventional differential privacy, the acceptable values of $\epsilon$ should be within a small range, ideally in the interval of ($0,9$]~\cite{abadi2016deep}. Due to the lower sensitivity of the interpolation process, increasing $\epsilon$ greater than $2$ may lower privacy. It is the users' responsibility to decrease or increase $\epsilon$ depending on the requirements. We suggest an $\epsilon$ of $1$ to have a balance between privacy and utility. If the user chooses an $\epsilon$ value less than $1$, the algorithm will provide higher randomization, hence providing higher privacy and lower utility, whereas lower privacy and a higher utility will be provided in case of an $\epsilon$ value higher than $1$. The selection of $ws$ depends specifically on the size
of the particular dataset. A comparably larger $ws$ can be chosen for a large dataset, while $ws$ can be smaller for a small dataset. For a static dataset,
$ws$ can range from a smaller value such as one-tenth the \textit{size of the dataset} to the \textit{full size of the dataset}. The minimum value of $ws$ should not go down to a small value (e.g. $<100$) because it increases the number of perturbation cycles and introduces an extreme level of randomization to the input dataset, resulting in poor utility. For a data stream, $ws$ is considered as the buffer size and can range from a smaller value to any number of tuples that fit in the memory of the computer. Further discussions on selecting suitable values for $\epsilon$ and $ws$ is provided in Section \ref{resdes}.

\begin{algorithm}[H]
	\caption{Steps of the perturbation algorithm: SEAL}\label{privatealgo}
	\begin{spacing}{1.2}
		
		\begin{algorithmic}[1]
			\Statex\textbf{Inputs} :
			
			\begin{tabular}{l c l} 
				$D               $ & $\gets $ & input dataset (numeric)\\
				$\epsilon $ & $ \gets $ & scale of Laplacian noise \\
				$ ws $ & $ \gets $ & data buffer/window size \\
				$ t $ & $ \gets $ & threshold for the maximum number of windows processed \\
				& &  before a data release (default value of $t=-1$)\\
			\end{tabular}
			\Statex\textbf{Outputs} :
			\begin{tabular}{l c l} 
				$D^p $ & $\ \gets $ & perturbed dataset
			\end{tabular}
			\State divide $D$ in to data partitions ($w_i$) of size $ws$
			\State $x=[1,\dots,ws]$
			\LState normalize $x$ within the bounds of $[0, 1]$
			\Foreach {\textbf{each} $w_i$}
			\LState rep=rep+1
			\LState $D^p=[]$ \Comment{empty matrix}
			\LState normalize data of $w_i$ within the bounds of $[0, 1]$
			\Foreach {\textbf{each} attribute $a_i$ \textbf{in} $w_i$}
			\LState $sa_i=sort(a_i)$ \Comment{sorted in ascending order} \label{step9}
			\LState generate $M$ \Comment{according to Equation \ref{matM}}
			\LState generate $B$ \Comment{according to Equation \ref{vectorB}}
			\LState $A=B*M^{-1}$ \label{linmat}
			\LState use $A=[a_1, a_2, a_3, a_4]$ to generate perturbed data ($a_i^p$) using $\Hat{f(x)}$ \Comment{refer Equation \ref{newapfunc}}
			\LState normalize $a_i^p$ within the bounds of $[0, 1]$ to generate $a_i^N$
			\LState normalize $a_i^N$ within the bounds of $[min(a_i), max(a_i)]$
			\LState resort $a_i^N$ to the original order of $a_i$ to generate $a_i^o$ \label{step16}
			\EndForEach 
			\LState merge all $a_i^o$ to generate $w_i^p$
			\LState $D^p=merge(D^p,w^p_i) $ 
			\If {rep==t}
            					
            \LState randomly swap the tuple of $D^p$ 
			\LState release $\ D^{p} $	
            \LState rep=0 \label{step23}
            \EndIf
			\EndForEach 
            \If {t==-1}
            \LState randomly swap the tuple of $D^p$ 
			\LState return $\ D^{p} $	
            \EndIf

			\Statex \textbf{End Algorithm}
		\end{algorithmic}
		
	\end{spacing}
\end{algorithm}

\subsection{A use case: SEAL integration in a healthcare smart cyber-physical system}

Biomedical and healthcare systems provide numerous opportunities and challenges for SCPS. Intelligent operating rooms and hospitals, image-guided surgery and therapy, fluid flow control for medicine and biological assays and the development of physical and neural prostheses are some of the examples for biomedical and healthcare systems which can be effectively facilitated and improved using SCPS~\cite{baheti2011cyber}. However, biomedicine and healthcare data can contain a large amount of sensitive, personal information. SEAL provides a practical solution and can impose privacy in such scenarios to limit potential privacy leak from such systems~\cite{baheti2011cyber}. 

\begin{figure}[H]
	\centering
	\scalebox{1}{
		\includegraphics[width=1\textwidth, trim=0cm 0cm 0cm
		0cm]{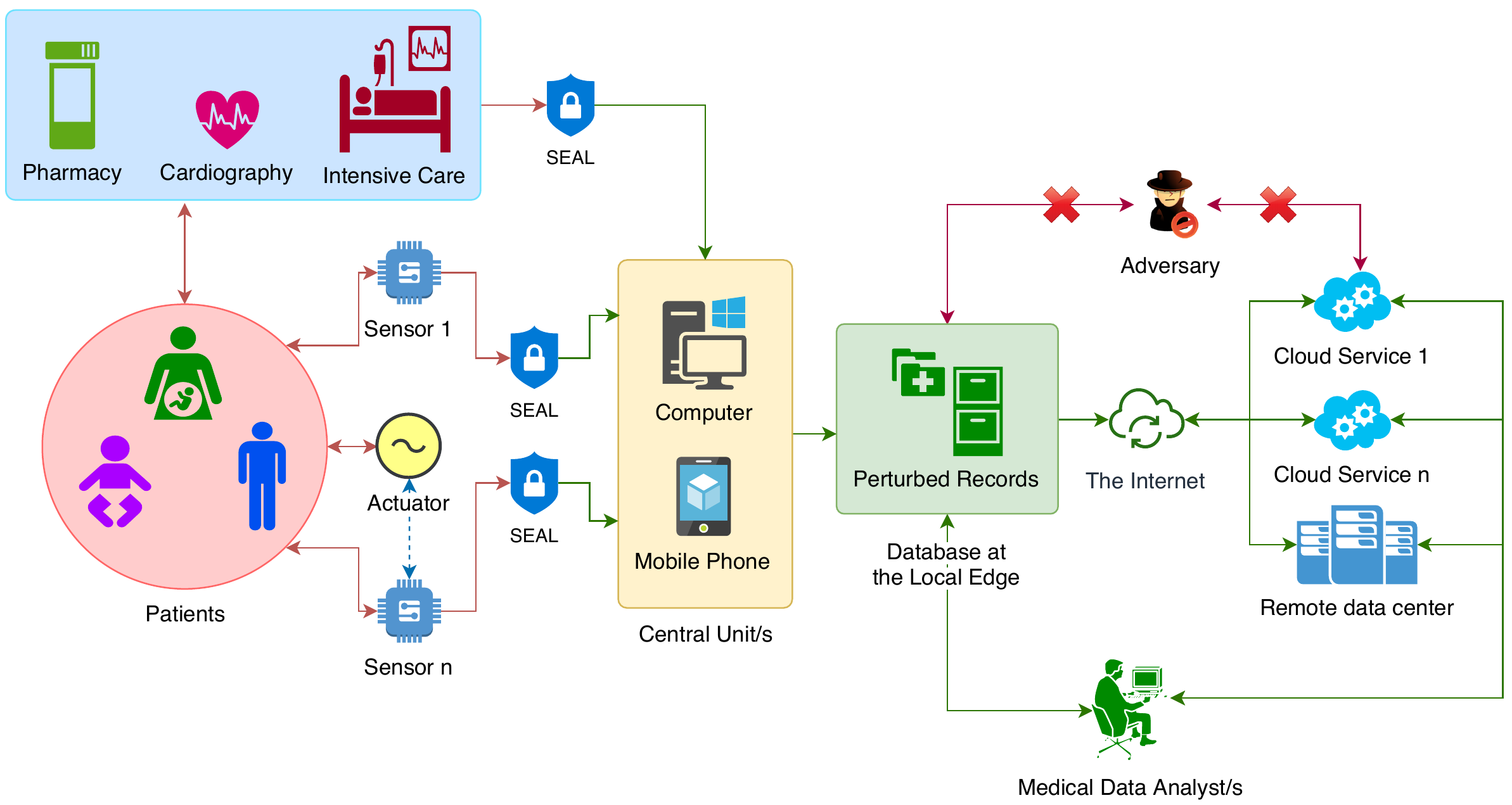}
	}
	\caption{A use case: The integration of SEAL in a healthcare smart cyber-physical system. As shown in the figure, SEAL perturbs data as soon as they leave the source (medical sensors, medical devices, etc.). In the proposed setting, SEAL assumes that there is no trusted party.}
	\label{sealusecase}
\end{figure}

Figure \ref{sealusecase} shows a  use case for SEAL integration in a healthcare smart cyber-physical system. Patients can have several sensors attached to them for recording different physical parameters. The
recorded data are then transmitted to a central unit which can be any readily available digital device such as a smartphone, a personal computer, or an embedded computer. A large variety of sensors are available today, e.g. glucose monitors, blood pressure monitors~\cite{alhayajneh2018biometric}. In the proposed setting, we assume that the processing unit that runs SEAL, perturbs all sensitive inputs forwarded to the central unit. As shown in the figure, we assume that the central units do not receive any unperturbed sensitive information, and the data repositories will store only perturbed data, locally or in a remote data center. Data analysts can access and use only the perturbed data to conduct their analyses. Since the data is perturbed, adversarial attacks on privacy will not be successful.   

\section{Experimental Results}
\label{expresults}

\begin{figure}[H]
	\centering
	\scalebox{1}{
		\includegraphics[width=0.4\textwidth, trim=0cm 0cm 0cm
		0cm]{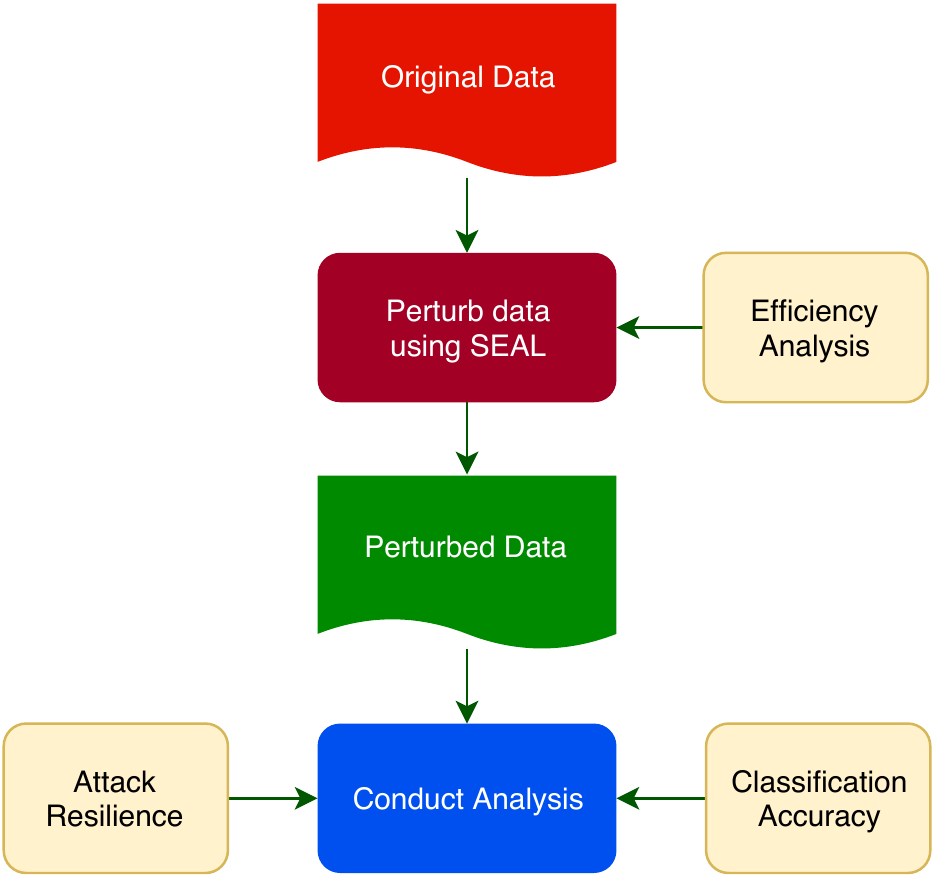}
	}
	\caption{The analytical setup used for SEAL. The figure shows the different levels of performance analysis of SEAL. First, the time consumption of perturbation was recorded under the efficiency analysis. Next, the attack resistance and classification accuracy were analyzed upon the perturbed data. }
	\label{sealanalytical}
\end{figure}

In this section, we discuss the experimental setup, resources used, experiments, and their results. The experiments were conducted using seven datasets retrieved from the UCI data repository\footnote{https://archive.ics.uci.edu/ml/index.php}. We compare the results of SEAL against the results of rotation perturbation (RP), geometric perturbation (GP) and data condensation (DC). For performance comparison with SEAL, we selected GP and RP when using static datasets, while DC was used with data streams. The main reason for selecting GP, RP, and DC is that they are multidimensional perturbation mechanisms that correlate with the technique used in the linear system of SEAL as given in Equation \ref{linsys}. Figure \ref{sealanalytical} shows the analytical setup which was used to test the performance of SEAL.  We perturbed the input data using SEAL, RP, GP, and DC and conducted data classification experiments on the perturbed data using five different classification algorithms to test and compare utility. We used the default settings of 10 iterations with a noise factor (sigma) of  0.3 to perturb the data using RP and GP.  Next, SEAL was tested for its attack resistance  against naive inference, known I/O attacks, and ICA-based attacks that are based on data reconstruction. These attacks are more successful against perturbation methods that use matrix multiplications. The attack resistance  results of SEAL were then compared with the attack resistance  results of RP, GP, and DC. Subsequently, we tested and compared SEAL's computational complexity and scalability by using two large datasets. Finally, we tested the performance of SEAL on data streams and compared the results with the results of DC.

\subsection{Experimental Setup}
\label{expsetup}

For the experiments we used a Windows 10 (Home 64-bit, Build 17134) computer with Intel (R) i5-6200U (6$^{th}$ generation) CPU (2 cores with 4 logical threads, 2.3 GHz with turbo up to 2.8 GHz) and 8192 MB RAM.  The scalability of the proposed algorithm was tested using a Linux (SUSE Enterprise Server 11 SP4) SGI UV3000 supercomputer, with 64 Intel Haswell 10-core processors, 25MB cache and 8TB of global shared memory connected by SGI's NUMAlink interconnect. We implemented SEAL in MATLAB R2016b. The experiments on data classification were carried out on Weka 3.6 \cite{witten2016data}, which is a collection of machine learning algorithms for data mining tasks.

\subsubsection{Datasets used in the experiments }
A brief description of the datasets is given in Table \ref{datasettb}. The dimensions of the datasets used for the performance testing vary from small to extremely large, to check the performance of SEAL in different circumstances.  Since the current version of SEAL conducts perturbation only on numerical data; we selected only numerical datasets in which the only attribute containing non-numerical data is the class attribute.

\begin{table}[H]
\centering

    \caption{Short descriptions of the datasets selected for testing.}   
    \label{datasettb}

    \begin{small}
    	\setlength\tabcolsep{5pt} 
        \resizebox{1\columnwidth}{!}{
    \begin{tabular}{l l l l l }
    \hline
{\bfseries Dataset} & {\bfseries  Abbreviation}         & {\bfseries Number of Records}     & {\bfseries Number of Attributes }   &   \bfseries{Number of Classes}   \\
    \hline
Wholesale customers\tablefootnote{https://archive.ics.uci.edu/ml/datasets/Wholesale+customers}       &	 WCDS & 440 \ & 8 \ & 2 	\\
    Wine Quality\tablefootnote{https://archive.ics.uci.edu/ml/datasets/Wine+Quality}      & WQDS & 4898 \ & 12 \ & 7 \\
     Page Blocks Classification \tablefootnote{https://archive.ics.uci.edu/ml/datasets/Page+Blocks+Classification}          & PBDS  & 5473 \ &  11 \ &   5  \\
Letter Recognition\tablefootnote{https://archive.ics.uci.edu/ml/datasets/Letter+Recognition}             &  LRDS	& 20000 & 17 & 26\\
       Statlog (Shuttle)\tablefootnote{https://archive.ics.uci.edu/ml/datasets/Statlog+\%28Shuttle\%29}        &  SSDS & 58000 \ & 9 \ & 7 \\

HEPMASS\tablefootnote{https://archive.ics.uci.edu/ml/datasets/HEPMASS\#}      & HPDS & 3310816  \ & 28 \ & 2 \\
HIGGS\tablefootnote{https://archive.ics.uci.edu/ml/datasets/HIGGS\#}      & HIDS & 11000000  \ & 28 \ & 2 \\
    \hline
    \end{tabular}
    }
    \end{small} 
\end{table}

\subsubsection{Perturbation methods used for comparison}

Random rotation perturbation (RP), geometric data perturbation (GP), and data condensation (DC)  are three types of matrix multiplicative perturbation approaches which are considered to provide high utility in classification and clustering~\cite{okkalioglu2015survey}. In RP, the original data matrix is multiplied using a random rotation matrix which has the properties of an orthogonal matrix. A rotational matrix $R$ follows the property of $R\times R^T= R^T\times R=I$, where $I$ is the identity matrix. The application of rotation is repeated until the algorithm converges at the desired level of privacy ~\cite{chen2005random}. In GP, a random translation matrix is added to the process of perturbation in order to enhance privacy. The method accompanies three components: rotation perturbation, translation perturbation, and distance perturbation ~\cite{chen2011geometric}. Due to the isometric nature of transformations, the perturbation process preserves the distance between the tuples, resulting in high utility for classification and clustering.  RP and GP can only be used for static datasets in their current setting, due to their recursive approach to deriving the optimal perturbation. DC is specifically introduced for data streams.  In DC, data are divided into multiple homogeneous groups of predefined size (accepted as user input) in such a way that the difference between the records in a particular group is minimal,  and a certain level of statistical information about different records is maintained. The sanitized data is generated using a uniform random distribution based on the eigenvectors which are generated using the eigendecomposition of the characteristic covariance matrices of each homogeneous group \cite{aggarwal2004condensation}. 

\subsubsection{Classification algorithms used in the experiments}
\label{classificationalgo}
Different classes of classification algorithms employ different classification strategies~\cite{lessmann2015benchmarking}. To investigate the performance of SEAL with diverse classification methods, we chose five different algorithms as the representative of different classes, namely: Multilayer Perceptron (MLP) ~\cite{witten2016data}, k-Nearest Neighbor (kNN) ~\cite{witten2016data}, Sequential Minimal Optimization (SMO) ~\cite{scholkopf1999advances}, Naive Bayes ~\cite{witten2016data}, and J48 ~\cite{quinlan1993c4}, and tested SEAL for its utility in  terms of classification accuracy. MLP  uses back-propagation to classify instances ~\cite{witten2016data}. kNN  is a non-parametric method used for classification ~\cite{witten2016data}. SMO is an implementation of John Platt's sequential minimal optimization algorithm for training a support vector classifier ~\cite{scholkopf1999advances}. Naive Bayes is a fast classification algorithm based on probabilistic classifiers ~\cite{witten2016data}. J48 is an implementation of the decision tree based classification algorithm ~\cite{witten2016data}.

\subsection{Performance Evaluation of SEAL}

We evaluated the performance of SEAL with regard to classification accuracy, attack resistance, time complexity, scalability, and also looked at data streams. First, we generated perturbed data using SEAL, RP, GP, and DC for the datasets: WCDS, WQDS, PBDS, LRDS, and SSDS (refer to Table \ref{datasettb}) under the corresponding settings. The perturbed data were then used to determine classification accuracy and attack resistance for each perturbed dataset. During the classification accuracy experiments, $k$ of k-nearest neighbor (kNN) classification algorithm was kept at $1$. The aggregated results were rated using the nonparametric statistical comparison test, Friedman's rank test, which is analogous to a standard one-way repeated-measures analysis of variance \cite{howell2016fundamental}. We recorded the statistical significance values, and the Friedman's mean ranks (FMR) returned by the rank test. The time consumption of SEAL was evaluated using runtime complexity analysis. We ran SEAL on two large-scale datasets, HPDS and HIDS, to test its scalability. Finally, the performance of SEAL was tested on data streams by running it on the LRDS dataset, and the results were compared with those produced by DC.

\subsubsection{Effect of randomization on the degree of privacy }

\label{resdes}
	
\begin{figure}[H] 
	
	\centering
	\subfloat[Effect of  perturbations by SEAL with $\epsilon=1$ on the same data series in two different instances.]{\includegraphics[width=0.48\textwidth, trim=0cm 0cm 0cm 0cm]{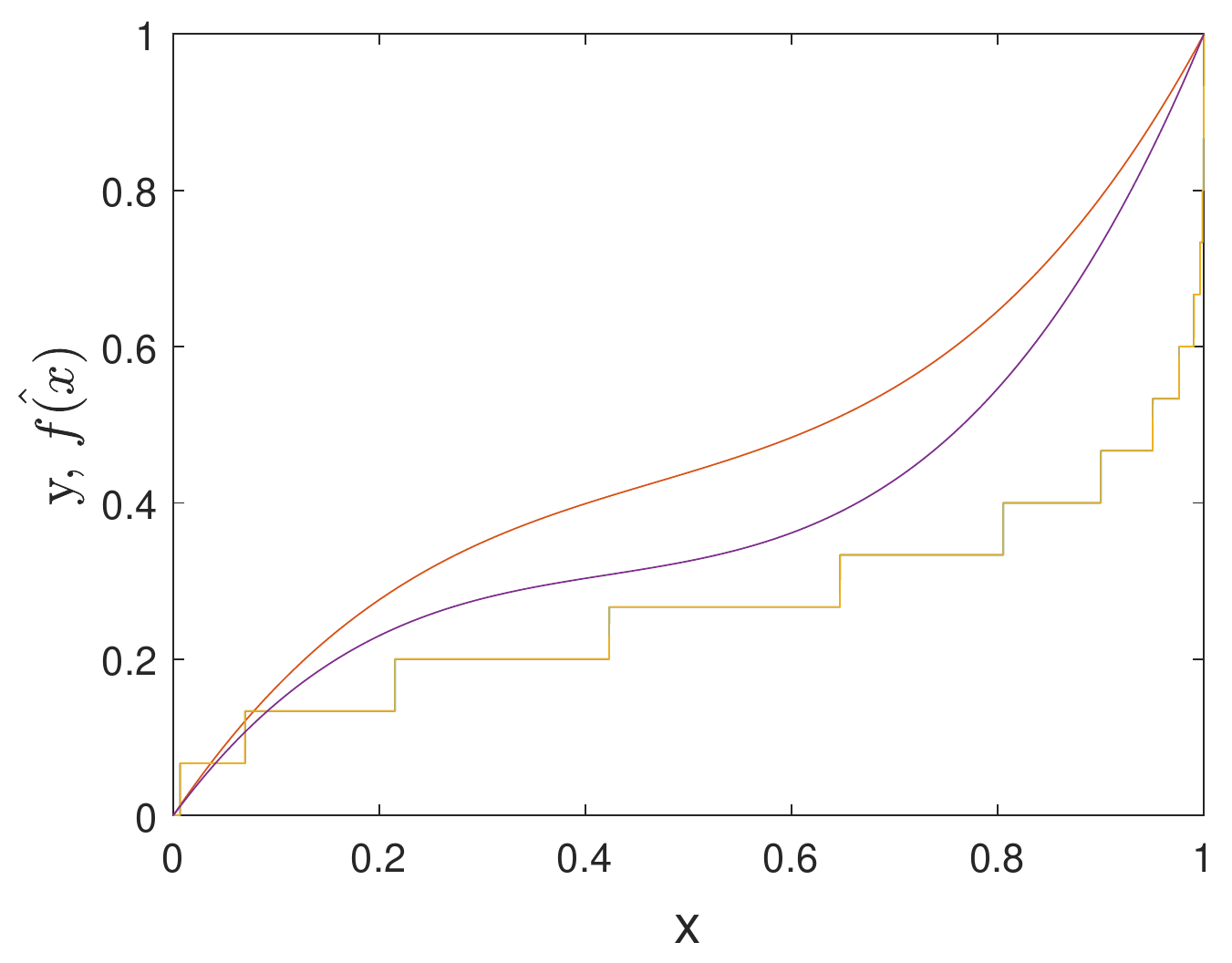}\label{chebpertinstepsilon1}}
	\hfill
	\subfloat[Effect of  perturbations by SEAL with $\epsilon=0.1$ on the same data series in two different instances.]{\includegraphics[width=0.48\textwidth, trim=0cm 0cm 0cm 0cm]{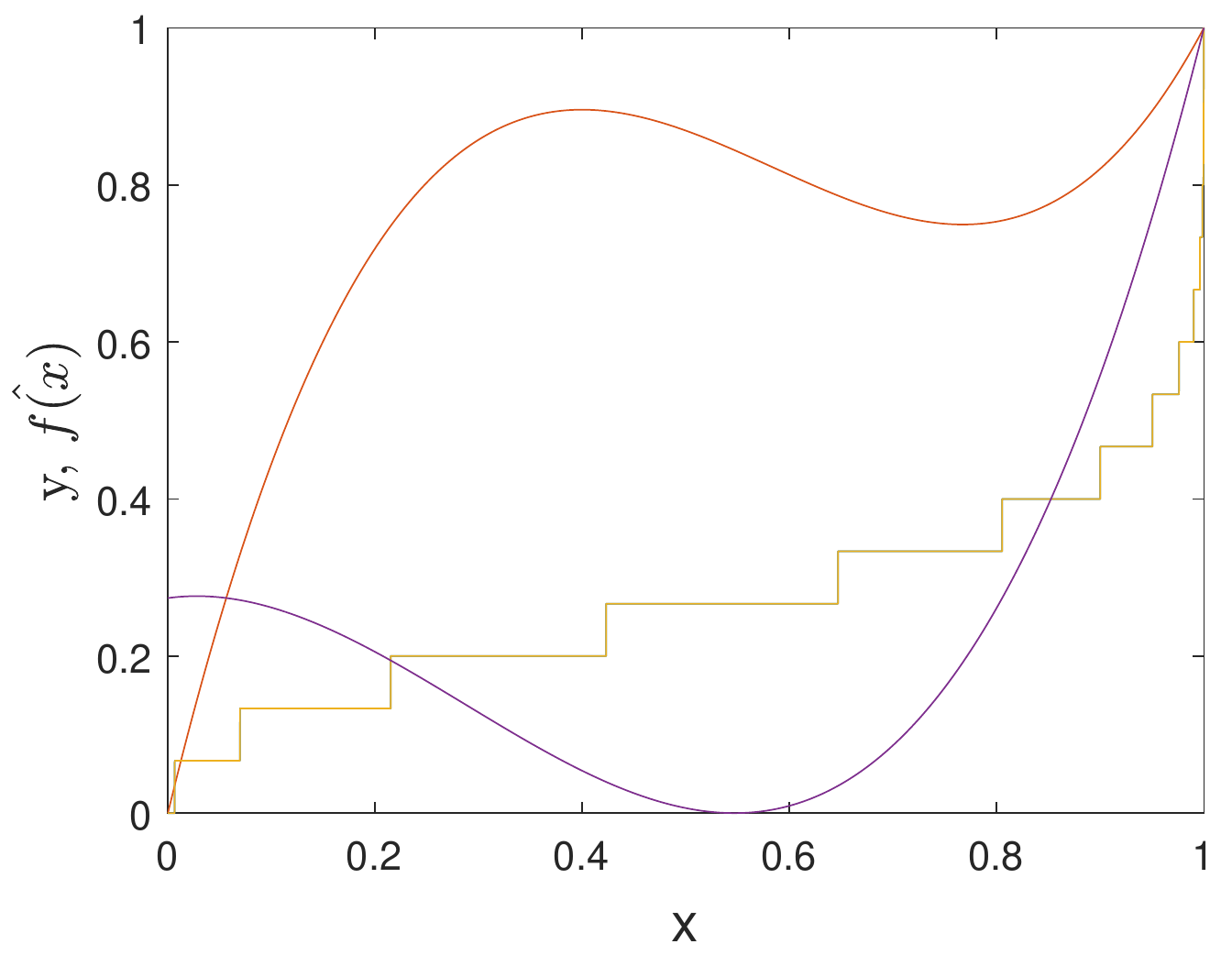}\label{chebpertinstepsilon01}}    
	\caption{Effect of perturbation by SEAL. The plot with the staircase pattern represents the original data series (first attribute of the LRDS dataset). The two plots that were plotted above the original data series represent two instances of perturbation conducted by SEAL on the original data series.}
    \label{chebpertinst}
    
    \medskip
\small
\end{figure}

One of the main features of SEAL is its ability to perturb a dataset while preserving the original shape of data distribution. We ran SEAL on the same data series to detect the effect of randomization in two different instances of perturbation. This experiment is to check and guarantee that SEAL does not publish similar perturbed data when it is applied with the same $\epsilon$ value to the same data on different occasions. This feature enables SEAL to prevent privacy leak via data linkage attacks that are exploiting multiple data releases. As depicted in Figure \ref{chebpertinst}, in two separate applications, SEAL generates two distinct randomized data series, while preserving the shape of the original data series. The left-hand plot of Figure \ref{chebpertinst} shows the data generated under an $\epsilon$ of $1$, whereas the right-hand plot shows the data generated under an $\epsilon$ of $0.1$. The right plot clearly shows the effect of high randomization under the extreme level of privacy generated by a strict privacy budget ($\epsilon$) of $0.1$. 

\subsubsection{Dynamics of  privacy budget ($\epsilon$) and window size ($ws$)}

\begin{figure}[H] 
	
	\centering
	\subfloat[Change of classification accuracy of the LRDS dataset perturbed by SEAL, where the window size ($ws$) was maintained at $10,000$ tuples. ]{\includegraphics[width=0.48\textwidth, trim=0cm 0cm 0cm 0cm]{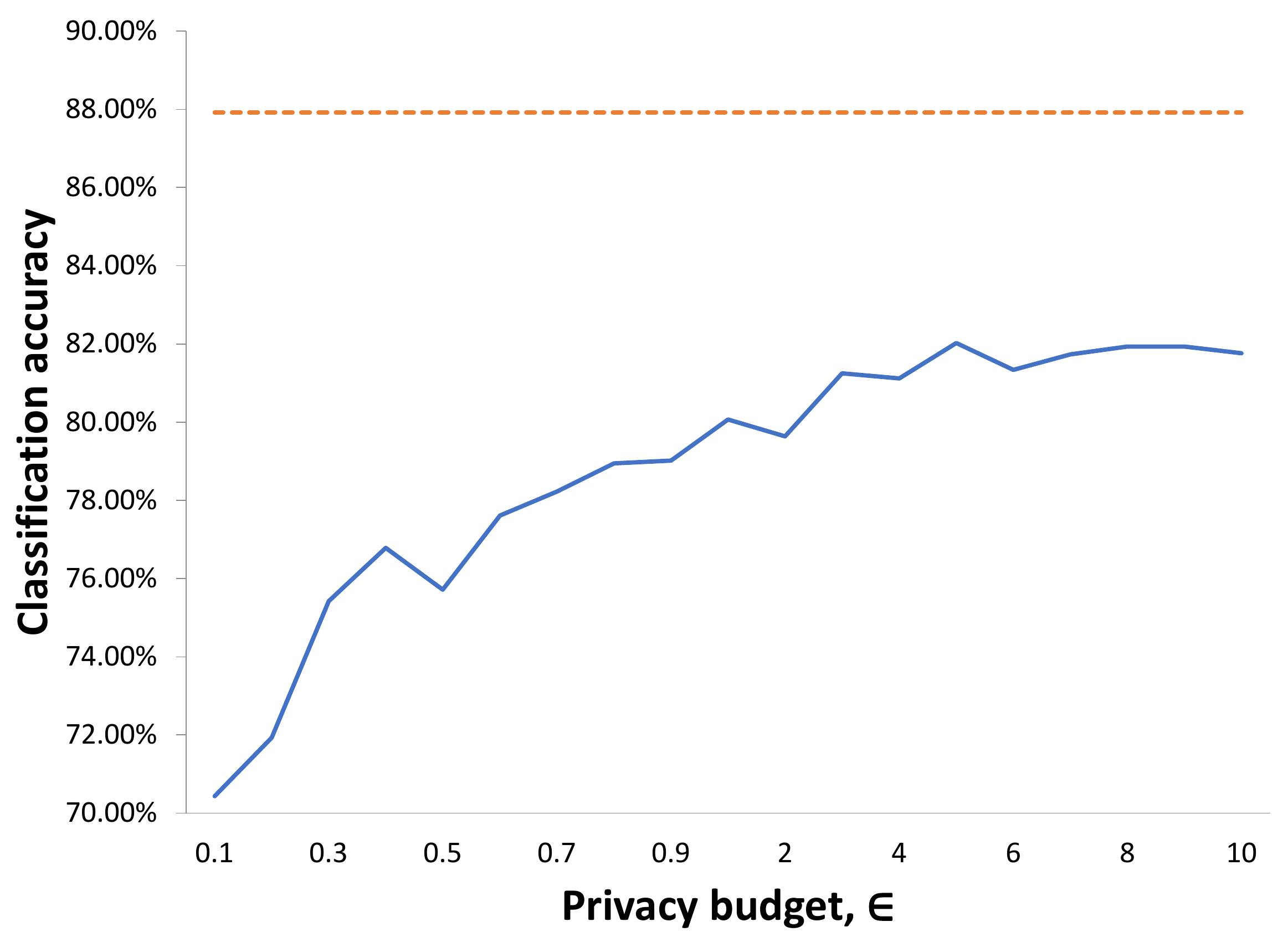}\label{changeclassaccu}}
	\hfill
	\subfloat[Change of classification accuracy (J48) of the LRDS dataset perturbed by SEAL where the privacy budget ($\epsilon$) was maintained at $1$. ]{\includegraphics[width=0.48\textwidth, trim=0cm 0cm 0cm 0cm]{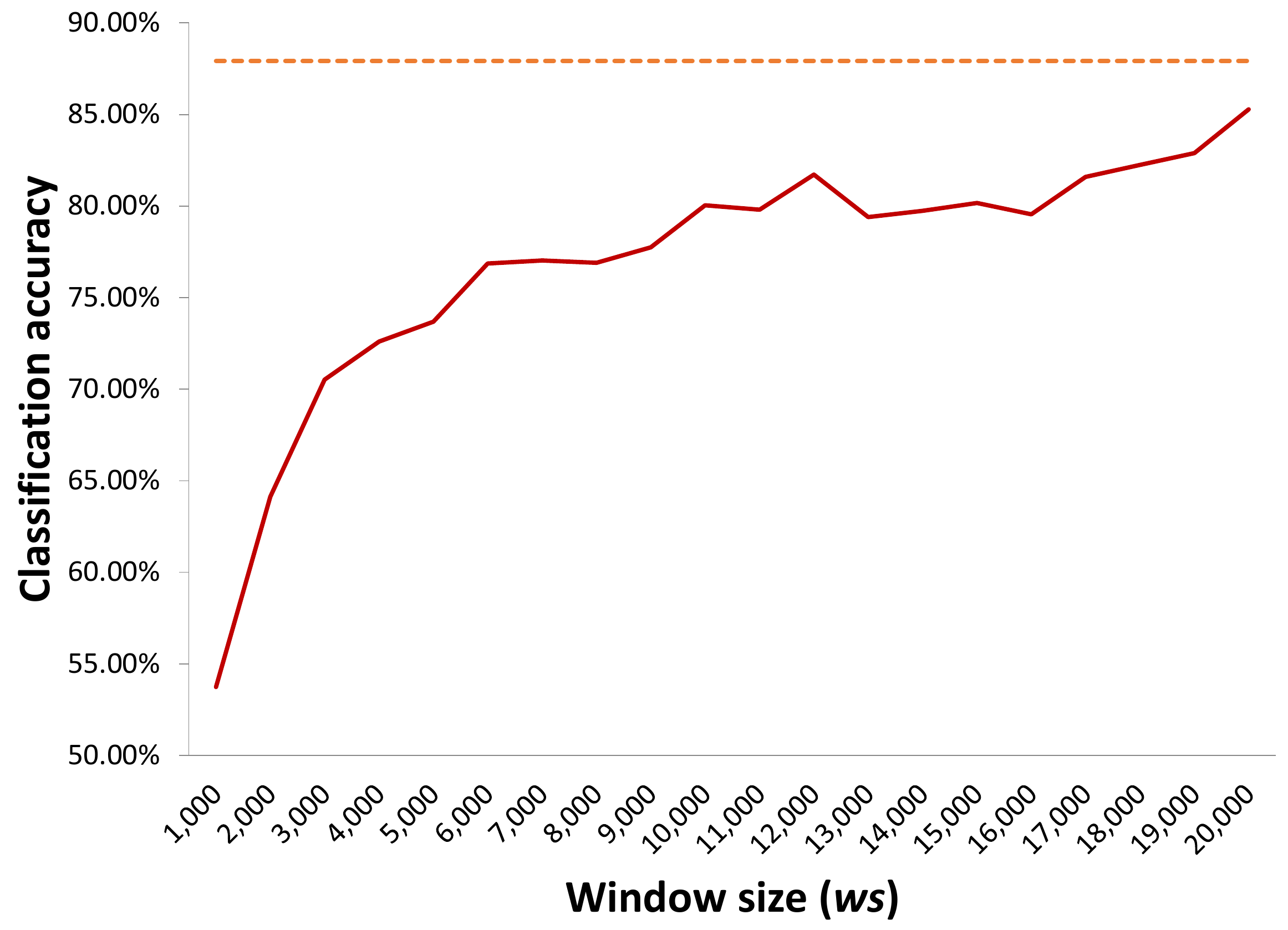}\label{changeclasswindow}}    
	\caption{Classification accuracy of SEAL. The classification accuracy was obtained by classifying the corresponding datasets using the J48 classification algorithm. The orange dotted horizontal lines on the two plots represent the classification accuracy of the original dataset. The window size ($ws$) is measured in number of tuples.}
    \label{changeclassaccufigure}
    
    \medskip
\small
\end{figure}

As explained in Section \ref{resdes}, smaller  $\epsilon$ means higher randomization, which results in decreased utility. Figure \ref{changeclassaccu} shows the change of classification accuracy against an increasing $\epsilon$. As shown in the figure, classification accuracy increases with an increasing privacy budget ($\epsilon$). Figure \ref{changeclassaccu} shows a more predictable pattern of increasing utility (classification accuracy) against increasing $\epsilon$. The choice of a proper $\epsilon$ depends
on the application requirements: a case that needs higher privacy should have a smaller $\epsilon$, while a larger  $\epsilon$ will provide better utility. As it turns out,  two-digit $\epsilon$ values provide no useful privacy. Given that SEAL tries to preserve the shape of the original data distribution, we recommend a range of $0.4$ to $3$ for
$\epsilon$ to limit unanticipated privacy leaks. We showed that SEAL provides better privacy and utility
than comparable methods under a privacy budget of $1$.  
 
Next, we tested the effect of window size ($ws$) on classification accuracy and the magnitude of randomization performed by SEAL. As shown on Figure \ref{changeclasswindow}, classification accuracy increases when $ws$ increases. When $ws$ is small, the dataset is divided into more groups than when $ws$ is large. When there is more than one group to be perturbed, SEAL applies randomization on each group distinctly. Since each of the groups is subjected to distinct randomization, the higher the number of groups, the larger the perturbation of the dataset. For smaller sizes of $ws$, SEAL will produce higher perturbation, resulting in more noise, reduced accuracy, improved privacy, and better resistance to data reconstruction attacks.

\subsubsection{Classification accuracy}

\begin{table}[htbp]
  \caption{Classification accuracies obtained when using the original dataset and the datasets perturned by three methods. During the experiments conducted using SEAL, the privacy budget $\epsilon$ was maintained at $1$. The window size $ws$ was maintained at full length of the corresponding dataset. For example, $ws$ of SEAL for the LRDS dataset was maintained at 20,000 during the experiments presented in this table. The last row shows the mean ranks returned by the nonparametric statistical comparison test (Friedman's rank test on the classification accuracies) of the three methods. A larger FMR value represents better classification accuracy.}  
  
   \label{classyaccuracy}
    \centering 
    \resizebox{0.8\columnwidth}{!}{
    \begin{tabular}{rllllll}
    \toprule
    \multicolumn{1}{l}{\textbf{Dataset}} & \textbf{Algorithm} & \multicolumn{1}{l}{\textbf{MLP}} & \multicolumn{1}{l}{\textbf{IBK}} & \multicolumn{1}{l}{\textbf{SVM}} & \multicolumn{1}{l}{\textbf{Naive Bayes}} & \multicolumn{1}{l}{\textbf{J48}} \\
    \midrule
    \multicolumn{1}{l}{LRDS} & Original   & 82.20\% & 95.96\% & 82.44\% & 64.01\% & 87.92\% \\
          & RP    & 74.04\% & 87.19\% & 71.07\% & 48.41\% & 64.89\% \\
          & GP    & 79.12\% & 93.05\% & 77.92\% & 59.89\% & 70.54\% \\
          & \textbf{SEAL} &  80.59\% &	93.67\% &	81.71\% &	63.10\% &	85.28\%\\
    \multicolumn{1}{l}{PBDS} & Original   & 96.25\% & 96.02\% & 92.93\% & 90.85\% & 96.88\% \\
          & RP    & 92.00\% & 95.52\% & 89.99\% & 35.76\% & 95.61\% \\
          & GP    & 90.24\% & 95.67\% & 89.93\% & 43.10\% & 95.49\% \\
          & \textbf{SEAL} &  96.34\% &	96.73\% &	95.59\% &	86.97\% &	96.34\%\\
    \multicolumn{1}{l}{SSDS} & Original   & 99.72\% & 99.94\% & 96.83\% & 91.84\% & 99.96\% \\
          & RP    & 96.26\% & 99.80\% & 88.21\% & 69.04\% & 99.51\% \\
          & GP    & 98.73\% & 99.81\% & 78.41\% & 79.18\% & 99.59\% \\
          & \textbf{SEAL} &  99.70\% &	99.21\% &	98.51\% &	89.94\% &	99.87\%\\
    \multicolumn{1}{l}{WCDS} & Original   & 90.91\% & 87.95\% & 87.73\% & 89.09\% & 90.23\% \\
          & RP    & 89.09\% & 85.00\% & 82.27\% & 84.55\% & 86.82\% \\
          & GP    & 91.82\% & 86.59\% & 85.00\% & 84.32\% & 88.86\% \\
          & \textbf{SEAL} &  89.32\% &	86.82\% &	89.09\% &	88.41\% &	86.59\%\\
    \multicolumn{1}{l}{WQDS} & Original   & 54.94\% & 64.54\% & 52.14\% & 44.67\% & 59.82\% \\
          & RP    & 47.65\% & 53.29\% & 44.88\% & 32.32\% & 45.53\% \\
          & GP    & 48.86\% & 56.88\% & 44.88\% & 32.16\% & 46.43\% \\
          & \textbf{SEAL} &  53.92\% &	64.02\% &	52.02\% &	47.83\% &	84.15\%\\
          \midrule
          \multicolumn{1}{c}{\bfseries FMR Values}& \multicolumn{2}{c}{RP: 1.34 }& \multicolumn{2}{c}{GP: 1.86}& \multicolumn{2}{c}{SEAL: 2.80}\\
    \bottomrule
    \end{tabular}%
    }

\medskip
\small
\end{table}%

Table \ref{classyaccuracy} provides the classification accuracies when using the original dataset and the datasets perturbed by the three methods. During the experiments for classification accuracy, we maintained $\epsilon$ at $1$ and $ws$ at the total length of the dataset. For example, if the dataset contained $n$ number of tuples, $ws$ was maintained at $n$. After producing the classification accuracies, Friedman's rank test was conducted on the data available in Table \ref{classyaccuracy} to rank the three methods: GP, RP, and SEAL. The mean ranks produced by Friedman's rank (FR) test are presented in the last row of Table \ref{classyaccuracy}\footnote{The FR test returned a $\chi^2$ value of 27.6566, a degree of freedom of 2 and a p-value of 9.8731e-07.}. The p-value suggests that the difference between the classification accuracies of RP, GP, and SEAL are significantly different. When evaluating FMR values on classification accuracies, a higher rank means that the corresponding method tends to produce better classification results. The mean ranks indicate that SEAL provides comparatively higher classification accuracy. SEAL is capable of providing higher utility in terms of classification accuracy due to its ability to maintain the shape of the original data distribution despite the introduced randomization. Although SEAL provides better performance overall than the other two methods, we can notice that in a few cases (as shown in Table \ref{classyaccuracy}) SEAL has produced slightly lower classification accuracies. We assume that this is due to the effect of variable random noise applied by SEAL. However, these lower accuracies are still on par with accuracies produced by the other two methods.

\subsubsection{Attack resistance}
Table \ref{attackresilience} shows the three methods' (RP, GP, and SEAL) resistance to three attack methods: naive snooping (NI), independent component analysis (ICA) and known I/O attack (IO)~\cite{chen2005random, okkalioglu2015survey}. We used the same parameter settings of SEAL ($\epsilon=1$ and $ws$=number of tuples) which were used in classification accuracy experiments for attack resistance analysis as well.  IO and ICA data reconstruction attacks try to restore the original data from the perturbed data and are more successful in attacking matrix multiplicative data perturbation.  FastICA package ~\cite{gavert2005fastica} was used to evaluate the effectiveness of ICA-based reconstruction of the perturbed data.  We obtained the attack resistance values as standard deviation values of (i)
the difference between the normalized original data and the perturbed data for NI, and (ii) the difference between the normalized original data and reconstructed data for ICA and IO. During the IO attack analysis, we assume that around 10\% of the original data is known to the adversary. The ``$min$" values under each test indicate the minimum guarantee of resistance while ``$avg$" values give an impression of the overall resistance.

We evaluated the data available in Table \ref{attackresilience}  using Friedman's rank test to generate the mean ranks for GP, RP, and SEAL. The mean ranks produced by Friedman's rank test are given in the last row of Table \ref{attackresilience}\footnote{The test statistics: $\chi^2$ value of 14.6387, a degree of freedom of 2 and a p-value of 6.6261e-04.}. The p-value implies that the difference between the attack resistance values is significantly different. As for the FMR values on attack resistance, a higher rank means that the corresponding method tends to be more attack-resistant. The mean ranks suggest that SEAL provides comparatively higher security than the comparable methods against the privacy attacks. 

\begin{table}[H]
	\centering
	
	\caption{Attack resistance  of the algorithms. During the experiments conducted using SEAL, the privacy budget $\epsilon$ was maintained at $1$. The window size $ws$ was maintained at full length of the corresponding dataset. For example, $ws$ of SEAL for the LRDS dataset was maintained at 20,000 during the experiments presented in this table. The last row provides Friedman's mean ranks returned by the nonparametric statistical comparison test on the three methods.}     
	\label{attackresilience}
	
	\begin{small}
    \resizebox{0.9\columnwidth}{!}{
		\begin{tabular}{l l l l l l l l l }
			\hline
			{\bfseries Dataset} & {\bfseries Algorithm} &  {\bfseries  $NI_{min}$ }         & {\bfseries $NI_{avg}$ }     & {\bfseries $ICA_{min}$  } & {\bfseries  $ICA_{avg}$ }  & {\bfseries $IO_{min}$ }     & {\bfseries $IO_{avg}$  } \\
			\hline
			 LRDS& RP & 0.8750&1.4490&0.4057&0.6942&0.0945&0.2932\\
			 & GP & 1.3248&1.6175&0.6402&0.7122&0.0584&0.4314\\
			 & SEAL & 1.4061  & 1.4148 & 0.7024 & 0.7062 & 0.6986 & 0.7056\\
			PBDS& RP & 0.7261&1.3368&0.5560&0.6769&0.0001&0.1242\\
			 & GP &0.2845&1.4885&0.1525&0.6834&0.0000&0.1048\\ 
			& SEAL & 1.3900  & 1.4084 & 0.7008 & 0.7056 & 0.6932 & 0.7031\\
			SSDS& RP & 1.2820&1.5015&0.1751&0.5909&0.0021&0.0242\\
			 & GP & 1.4490&1.6285&0.0062&0.3240&0.0011&0.0111\\
			& SEAL & 1.4065 & 1.4119 & 0.7038 & 0.7068 & 0.7027 & 0.7068\\
			WCDS & RP &1.0105&1.3098&0.6315&0.7362&0.0000&0.0895 \\
			& GP & 1.4620&1.7489&0.1069&0.6052&0.0000&0.1003\\
			& SEAL & 1.3130  & 1.3733 & 0.6775 & 0.7053 & 0.6557 & 0.6930\\
			WQDS& RP & 1.2014&1.4957&0.4880&0.7062&0.0057&0.4809 \\
			 & GP & 1.3463&1.6097&0.3630&0.6536&0.0039&0.4025\\
            & SEAL & 1.3834  & 1.4138 & 0.7018 & 0.7053 & 0.6859 & 0.7026\\
			\midrule
			\multicolumn{2}{c}{\bfseries FMR Values}& \multicolumn{2}{c}{RP: 1.68 }& \multicolumn{2}{c}{GP: 1.75}& \multicolumn{2}{c}{SEAL: 2.57 }\\
            \hline
		\end{tabular}
        }
	
	\end{small} 
\medskip
\small
	
\end{table}

\subsubsection{Time complexity}

\begin{figure}[H] 
	
	\centering
	\subfloat[Change of the time elapsed for the LRDS dataset with an increasing number of instances. ]{\includegraphics[width=0.48\textwidth, trim=0cm 0cm 0cm 0cm]{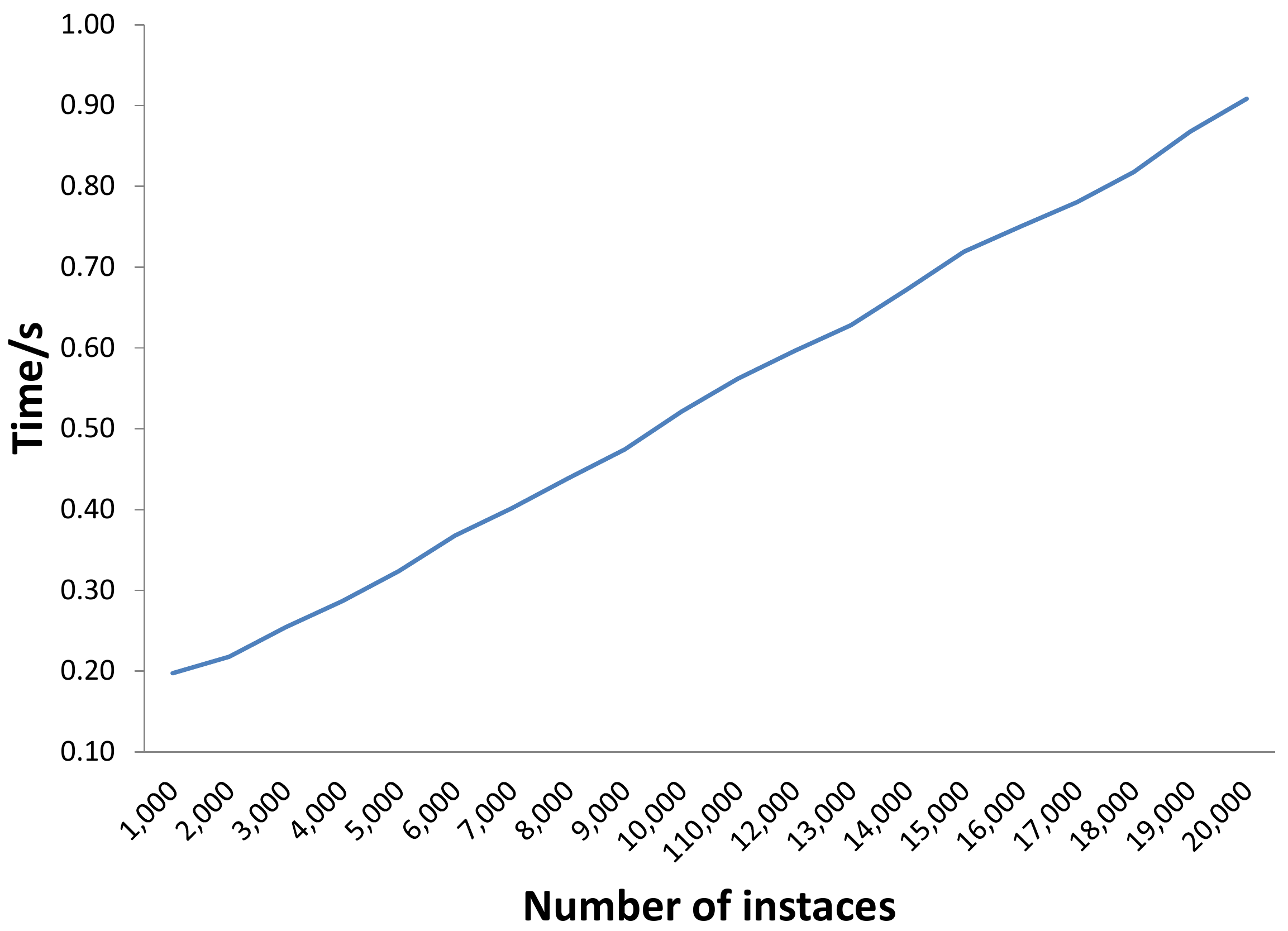}\label{timecomparisoninstances}}
	\hfill
	\subfloat[Change of the time elapsed for the LRDS dataset with an increasing number of attributes.]{\includegraphics[width=0.48\textwidth, trim=0cm 0cm 0cm 0cm]{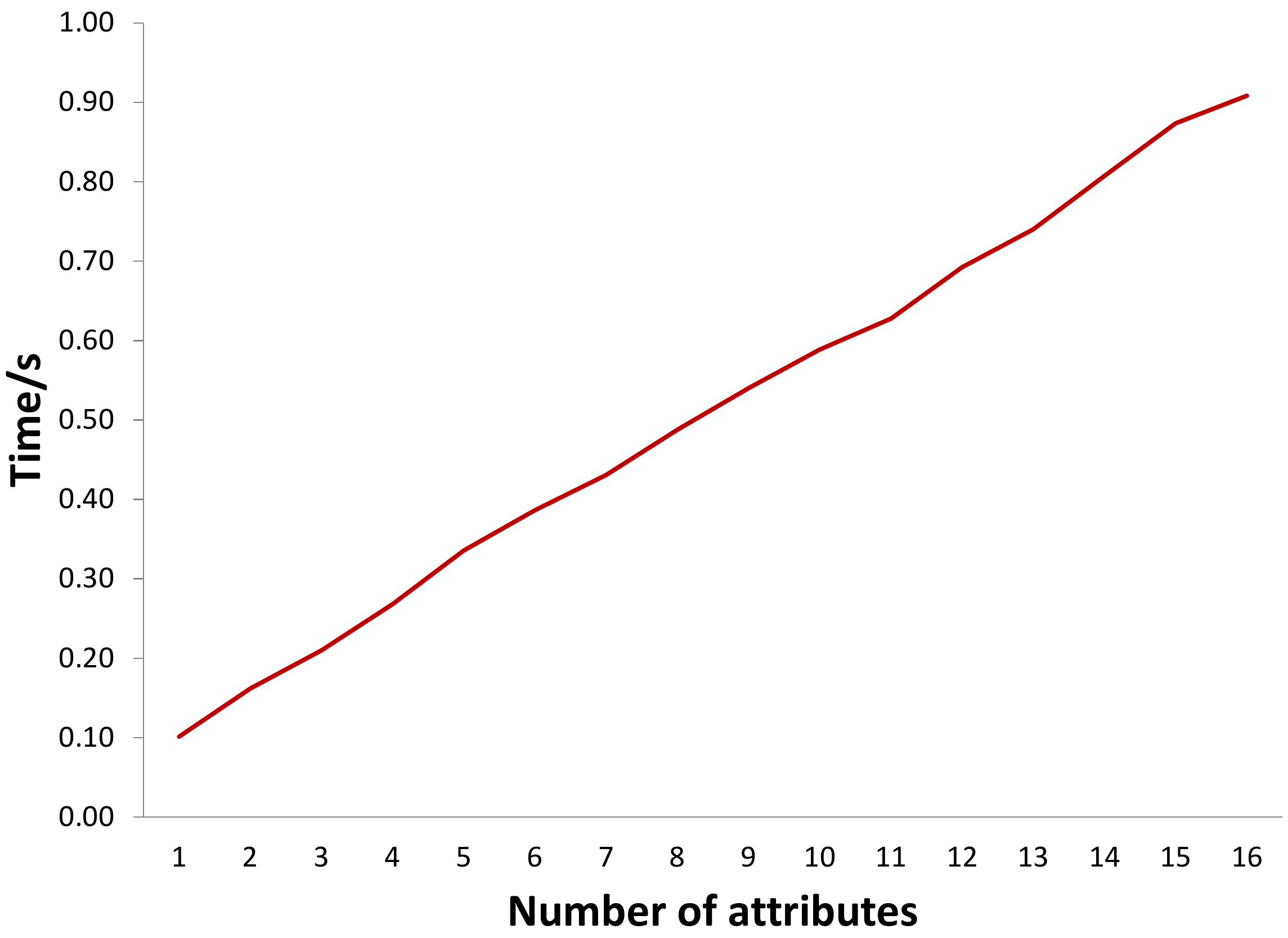}\label{timecomparison}}    
	\caption{Time consumption of SEAL.  During the runtime analysis, the window size ($ws$) was maintained at a full length of the corresponding instance of LRDS. The privacy budget ($\epsilon$) was maintained at $1$.}
    \label{timecompthree}
    \medskip
    \small
\end{figure}

Algorithm \ref{privatealgo} (SEAL) has two loops. One loop is controlled by the number of data partitions resulting from the window size ($ws$), and the number of attributes controls the other loop. In a particular instance of perturbation, these two parameters ($ws$ and the number of attributes) remain constants. Let us take the contribution of both loops to the computational complexity as a constant ($k$). If we evaluate the steps of SEAL from step \ref{step9} to step \ref{step23}, we can see that the highest computational complexity in these steps is $O(n)$, where $n$ is the number of tuples. From this, we can estimate the time complexity of Algorithm \ref{privatealgo}  to be $O(kn) = O(n)$. We investigated the time consumption against the number of instances and the number of attributes to determine the computational complexity empirically. Figure \ref{timecompthree} confirms that the time complexity of SEAL is in fact $O(n) $.  

\subsubsection{Time complexity comparison}

\begin{figure}[H] 
	
	\centering
	\subfloat[Increase of time consumption of SEAL, RP, and GP against the number of tuples.]{\includegraphics[width=0.48\textwidth, trim=0cm 0cm 0cm 0cm]{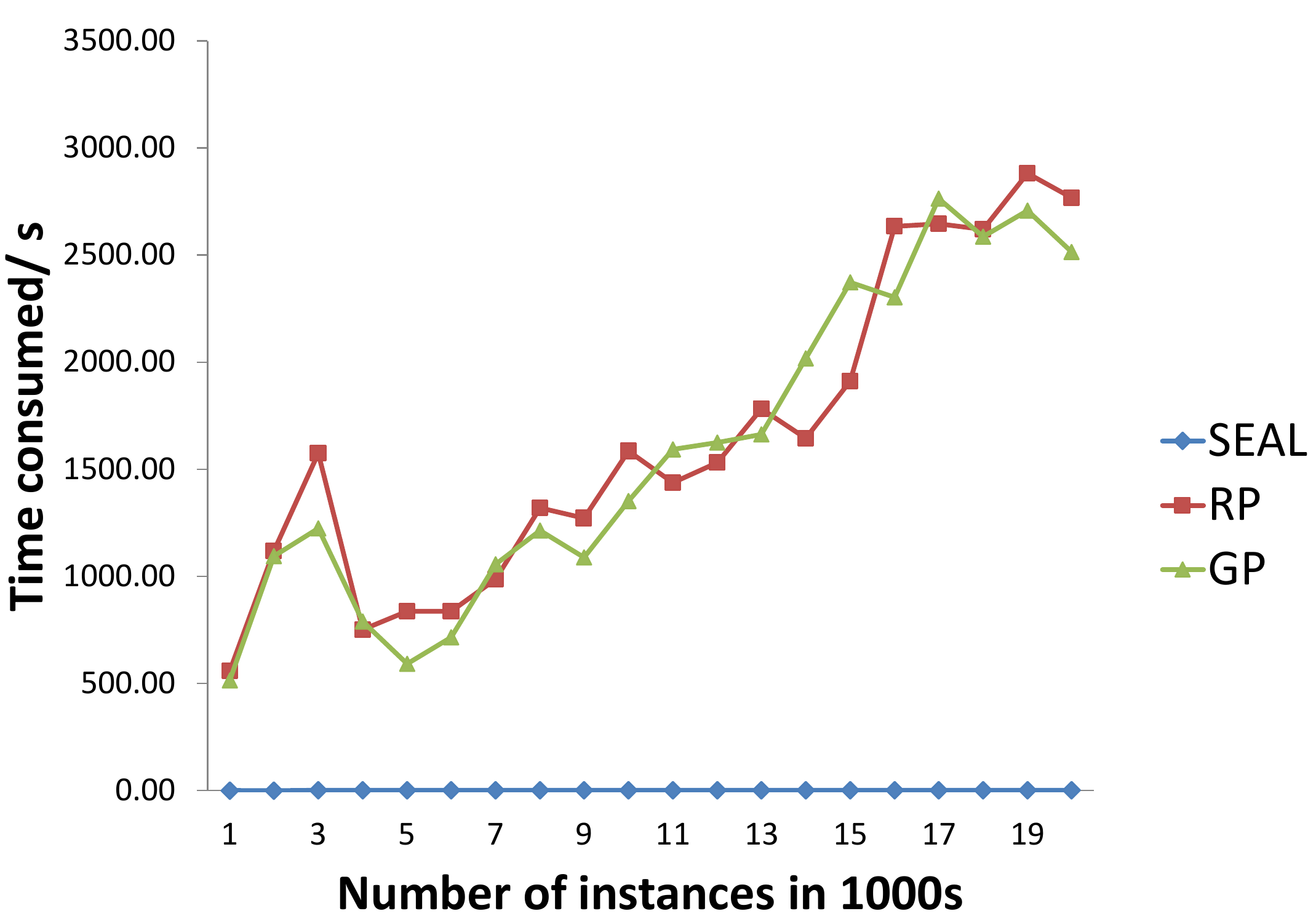}\label{timecomparisoninstances2}}
	\hfill
	\subfloat[Time consumption of SEAL, RP, and GP against the number of attributes.]{\includegraphics[width=0.48\textwidth, trim=0cm 0cm 0cm 0cm]{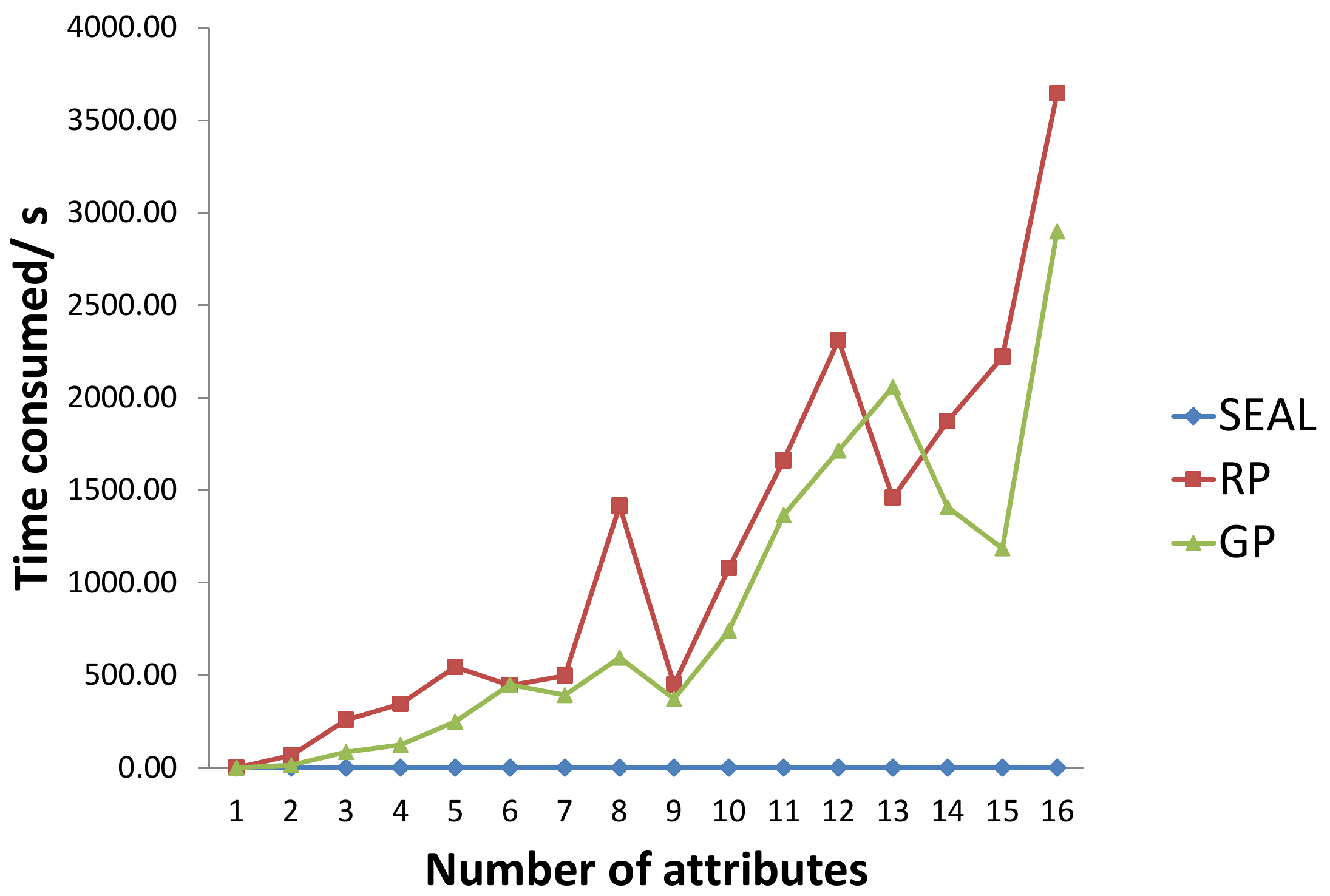}\label{timecomparison2}}    
	\caption{Time consumption comparison of SEAL, RP, and GP. The time consumption plots available in Figure \ref{timecompthree} are plotted in comparison with the time consumption plots of RP and GP. Due to the extremely low time consumption of SEAL, its curves lie almost on the x-axis when drawn in a plot together with the others.}
	

    \label{seal_timecomp}
 
\end{figure}

Both RP and GP show $O(n^2)$ time complexity to perturb one record with $n$ attributes. The total complexity to perturb a dataset of $m$ records is $O(m\times n^2)$. However, both RP and GP run for $r$  number of iterations (which is taken as a user input) to find the optimal perturbation instance of the dataset within the $r$ iterations. Therefore, the overall complexity is $O(m\times r \times n^2)$. Under each iteration of $r$, the algorithms run data reconstruction using ICA and known IO attacks to find the vulnerability level of the perturbed dataset.  Each attack runs another $k$ number of iterations (which is another user input) to reconstruct $k$ number of instances. Usually, $k$ is much larger than $r$. For one iteration of $k$, IO and ICA contribute a complexity of $O(n\times m)$~\cite{zarzoso2006fast}. Hence, the overall complexity of RP or GP in producing an optimal perturbed dataset is equal to $O(m^2\times r \times k \times n^3)$ which is a much larger computational complexity compared to the linear computational complexity of SEAL. Figure \ref{seal_timecomp} shows the time consumption plots of the three methods plotted together on the same figure. As shown on the figures, the curves of SEAL  lie almost on the x-axis due to its extremely low time consumption compared to the other two methods.

\subsubsection{Scalability}

We conducted the scalability analysis of SEAL on an SGI UV3000 supercomputer (a detailed specification of the supercomputer is given in Section \ref{expsetup}). SEAL was tested for its scalability on two large datasets: HPDS and HIDS. The results are given in Table \ref{scalability}. It is apparent that SEAL is more efficient than RP, GP, and DC; in fact, RP and GP did not even converge after 100 hours (the time limit of the batch scripts were set to 100 h). Both RP and GP use recursive loops to achieve optimal perturbation, which slows down the perturbation process. Therefore, RP and GP are not suitable for perturbing big data and data streams. DC is effective in perturbing big data, but SEAL performs better by providing better efficiency and utility. 

\begin{table}[H]
	\centering
	\caption{Scalability results (in seconds) of the three methods for high dimensional data. }
	\resizebox{0.8\columnwidth}{!}{
	\begin{tabular}{l l r r r}
		\toprule
		\textbf{Dataset} & \textbf{RP} & \multicolumn{1}{l}{\textbf{GP}} & \multicolumn{1}{l}{\textbf{DC (k=10,000)}} & \multicolumn{1}{l}{\textbf{SEAL ($ws$=10,000)}} \\
		\midrule
		HPDS  & NC within 100h & NC within 100h & 526.1168 & 97.8238 \\
		HIDS  & NC within 100h & NC within 100h & 6.42E+03 &  1.02E+03 \\
		\bottomrule
	\end{tabular}%
}

\small
\footnote{Footnote}NC: Did not converge
	\label{scalability}%
\end{table}%

\subsubsection{Performance on data streams}

\begin{figure}[H]
	\centering
	\subfloat[Change of classification accuracy against the buffer size.]{\includegraphics[width=0.49\textwidth, trim=0cm 0cm 0cm 0cm]{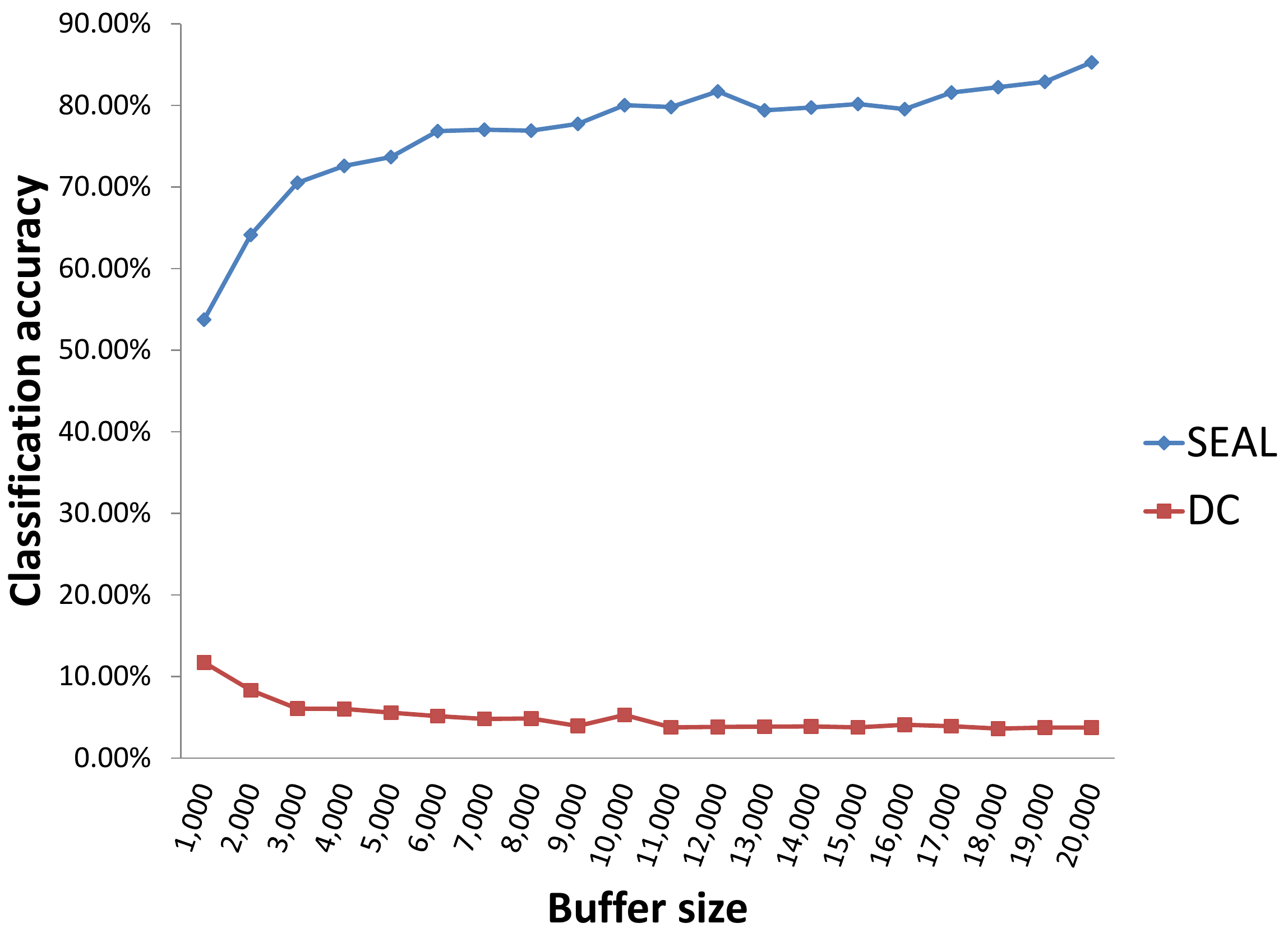}\label{classacbfsize}}
	\hfill
	\subfloat[Change of $minimum$ $STD(D-D^p)$ consumption against the buffer size.]{\includegraphics[width=0.49\textwidth, trim=0.3cm 0cm 0cm 0cm]{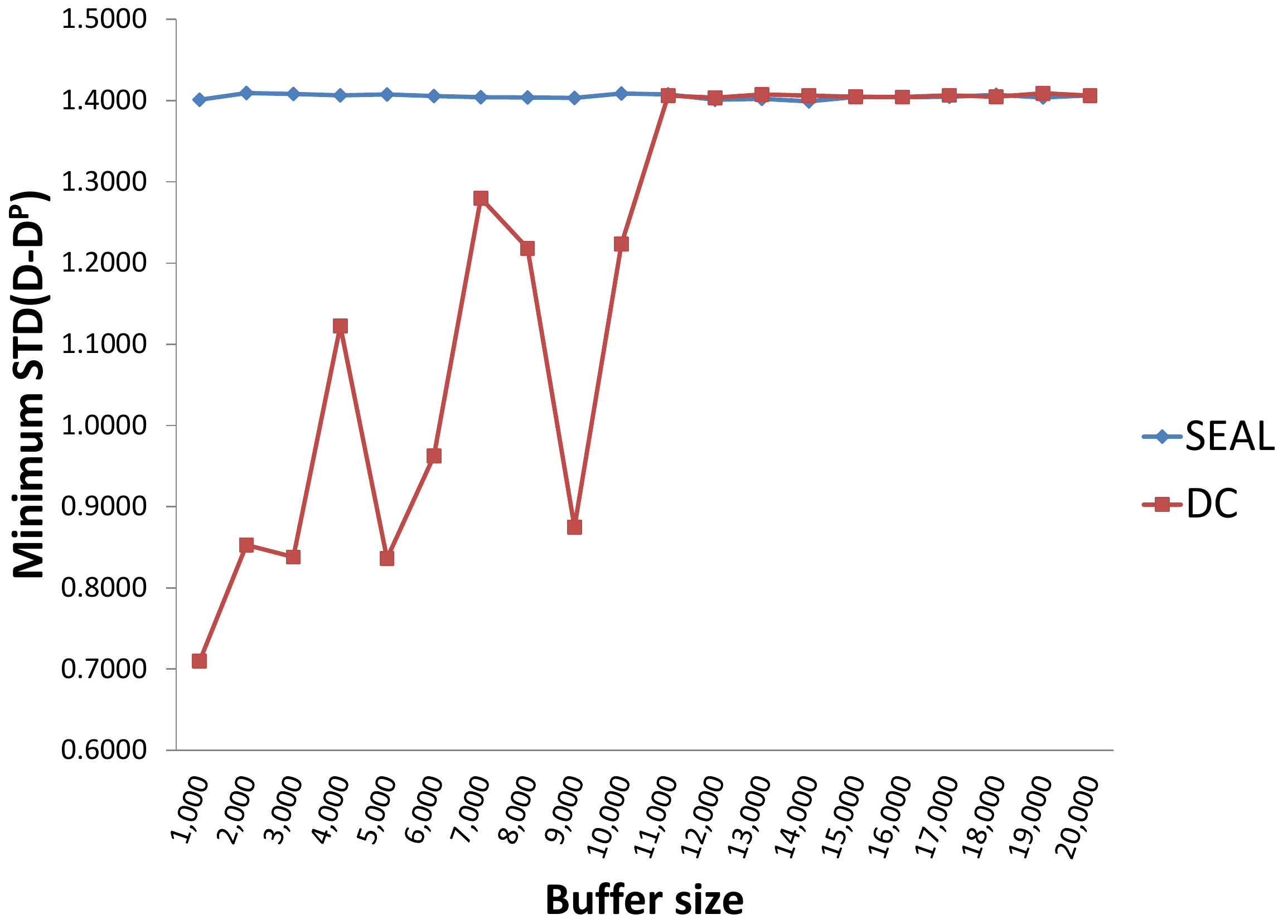}\label{minstdbfsize}}
 	\caption{Dynamics of classification accuracy and the minimum STD(D-D$^p$) against increasing  buffer size. During the experiments, the privacy budget ($\epsilon$) was maintained at $1$. The $minimum$ $STD(D-D^p)$ represents the minimum (attribute) value of the standard deviation of the difference between the normalized attributes of the original data (LRDS dataset) and the perturbed data. The classification accuracy was obtained by classifying each perturbed datasets using the J48 classification algorithm.}
 	\label{streamdynamics}
 	\medskip
    \small
\end{figure}

We checked the performance of SEAL on data streams with regard to (i) classification accuracy and (ii) $Minimum$ $STD(D-D^p)$. 
The latter provides evidence to the minimum guarantee of attack resistance provided under a particular instance of perturbation. As shown in Figure \ref{classacbfsize}, the classification accuracy of SEAL increases with increasing buffer size. This property is valuable for the perturbation of infinitely growing data streams generated by systems such as smart cyber-physical systems. The figure indicates that when a data stream grows infinitely, the use of smaller window sizes would negatively affect the utility of the perturbed data. When the window size is large, the utility of the perturbed data is closer to the utility of the original data stream. We can also notice that DC performs poorly in terms of classification accuracy compared to SEAL. It was previously noticed that DC works well only for tiny buffer sizes such as $5$ or $10$~\cite{chamikaraprocal}. However, according to Figure \ref{minstdbfsize}, the minimum guarantee of attack resistance drops when the buffer size decreases, which restricts the use of DC with smaller buffer sizes. According to Figure \ref{minstdbfsize}, however, SEAL still provides a consistent minimum guarantee of attack resistance, which allows SEAL to be used with any suitable buffer size.

\section{Discussion}
\label{discussion}
The proposed privacy-preserving
mechanism (named SEAL) for big data and data streams performs data perturbation based on Chebyshev polynomial interpolation and the application of a Laplacian mechanism for noise addition. SEAL uses the first four orders of Chebyshev polynomials of the first kind for the polynomial interpolation of a particular dataset. Although Legendre polynomials would offer a better approximation of the original data during interpolation,  Chebyshev polynomials are simpler to calculate and provide improved privacy; a higher interpolation error, i.e. increased deviation from the original data would intuitively provide greater privacy than Legendre polynomials. Moreover, we intend to maintain the spatial arrangement of the original data, and this requirement is fully satisfied by  Chebyshev interpolation.   During the interpolation, SEAL adds calibrated noise using the Laplacian mechanism to introduce randomization, and henceforth privacy, to the perturbed data.  The Laplacian noise allows the interpolation process to be performed with an anticipated random error for the root mean squared error minimization.  We follow the conventions of differential privacy for noise addition, the introduction of noise is in accordance with the characteristic privacy budget $\epsilon$. The privacy budget ($\epsilon$) allows users (data curators) of SEAL  to adjust the amount of noise. Smaller values of $\epsilon$ (usually less than 1 but greater than 0) add more noise to generate more randomization, whereas large values of $\epsilon$ add less noise and generate less randomization. The privacy budget is especially useful for multiple data release, where the data curator can apply proper noise in the perturbation process in consecutive data releases. SEAL's ability to maintain the shape of the original data distribution after noise addition is a clear advantage, and enables SEAL to provide convincingly higher utility than a standard local differentially private algorithm. This characteristic may come at a price, and the privacy enforced by a standard differentially private mechanism can be a little higher than that of SEAL.

The experimental results of SEAL show that it performs well on both static data and data streams. We evaluated SEAL in terms of classification accuracy, attack resistance, time complexity, scalability, and data stream performance. We tested each of these parameters using seven datasets, five classification algorithms, and three attack methods. SEAL outperforms the comparable methods: RP, GP, and DC in all these areas, proving that SEAL is an excellent choice for privacy preservation of data produced by SCPS and related technologies. SEAL produces high utility perturbed data in terms of classification accuracy, due to its ability to preserve the underlying characteristics such as the shape of the original data distribution. Although we apply an extensive amount of noise by using a small $\epsilon$ value, SEAL still tries to maintain the shape of the original data. The experiments show that even in extremely noisy perturbation environments, SEAL can provide higher utility compared to similar perturbation mechanisms, as shown in  Section \ref{expsetup}. SEAL shows excellent resistance with regard to data reconstruction attacks, proving that it offers excellent privacy. SEAL takes several steps to enhance the privacy of the perturbed data, namely (1) approximation through noisy interpolation, (2) scaling/normalization, and (3) data shuffling. These three steps help it outperform the other, similar perturbation mechanisms in terms of privacy.

In Section \ref{expsetup} we showed that SEAL has linear time complexity, $O(n)$. This characteristic is crucial for big data and data streams. The scalability experiments confirm that SEAL processes big datasets and data streams very efficiently. As shown in Figure \ref{streamdynamics}, SEAL also offers significantly better utility and attack resistance than data condensation. The amount of time spent by SEAL in processing one data record is around 0.03 to 0.09 milliseconds, which means that SEAL can perturb approximately 11110 to 33330 records per second. We note that runtime speed depends on the computing environment, such as CPU speed,  memory speed, and disk IO speeds. The processing speed of SEAL in our experimental setup suits many practical examples of data streams, e.g. Sense your City (CITY)\footnote{Sense your City  is an urban environmental monitoring project that used crowd-sourcing to deploy sensors at 7 cities across 3 continents in 2015 with about 12 sensors per city, and it generates 7000 messages/ sec.} and  NYC Taxi cab (TAXI)\footnote{NYC Taxi cab (TAXI)  produces a stream of smart transportation messages at the rate of 4000 messages/sec. The messages arrive from 2M trips taken on 20,355 New York city taxis equipped with GPS in 2013. }~\cite{shukla2016benchmarking}.    The results clearly demonstrate that SEAL is an efficient and reliable privacy preserving mechanism for practical big data and data stream scenarios.

\section{Conclusion}
\label{conclusion}
In this paper, we proposed a solution for maintaining data privacy in large-scale data publishing and analysis scenarios, which is becoming an important issue in various environments, such as smart cyber-physical systems. We proposed a novel algorithm named SEAL to perturb data to maintain data privacy. Linear time complexity  ($O(n)$) of SEAL allows it to work efficiently with continuously growing data streams and big data. Our experiments and comparisons indicate that SEAL produces higher classification accuracy, efficiency, and scalability while preserving better privacy with higher attack resistance than similar methods. The results prove that SEAL suits the dynamic environments presented by smart cyber-physical environments very well. SEAL can be an effective
privacy-preserving mechanism for smart cyber-physical systems such as vehicles, grid, healthcare systems, and homes, as it can effectively perturb continuous data streams generated by sensors monitoring an individual or group of individuals and process them on the edge/fog
devices before transmission to cloud systems for further analysis. 

The current configuration of SEAL does not allow distributed data perturbation, and it limits the utility only to privacy-preserving data classification. A potential future extension of SEAL can address a distributed perturbation scenario that would allow SEAL to perturb sensor outputs individually while capturing the distinct latencies introduced by the sensors. SEAL could then combine the individually perturbed data using the corresponding timestamps and latencies to produce the privacy-protected data records. Further investigation on privacy parameter
tuning would allow extended utility towards other areas such as descriptive statistics.

\appendix

\section{Chebyshev Polynomials of the First Kind}
\label{chebyshevapp}
\begin{mydef}
\label{defp1}
The Chebyshev polynomial $T_n(x)$ of the first kind is a polynomial in $x$ of degree $n$, defined by the relation,
\end{mydef}

\begin{equation}
    T_n(x)=cos \: n\theta \text{ \; when \; } x=cos \: \theta
    \label{cheb1}
\end{equation}

 From Equation \ref{cheb1}, we can deduce the first five ($n = 0,1,2,3,4$) Chebyshev polynomials using Equation \ref{seq1} to Equation \ref{seq5}, which are normalized such that $T_n(1)=1$, and  $x \in \left[-1, 1\right]$.

\begin{align}
T_0(x)=1\label{seq1}\\
T_1(x)=x\\
T_2(x)=2x^2-1\\
T_3(x)=4x^3-3x\\
T_4(x)=8x^4-8x^2+1\label{seq5}
\end{align}

Furthermore, we can represent any Chebyshev polynomial of the first kind using the recurrence relation given in Equation \ref{chebrecurse}, where $T_0(x)=1$ and $T_1(x)=x$.

\begin{equation}
T_{n+1}(x)=2xT_n(x)-T_{n-1}(x)
\label{chebrecurse}
\end{equation}

\section{Least Square Fitting}
\label{lsfapp}
In least square fitting, vertical least squares fitting proceeds by finding the sum of squares of the vertical derivations $R^2$ (refer Equation \ref{vlsf}) of a set of $n$ data points~\cite{weisstein2002least}.

\begin{equation}
    R^2\equiv \sum \left[ f(x_i,a_1, a_2, \dots, a_n) - y_i \right]^2
    \label{vlsf}
\end{equation}

Now, we can choose to minimize the quantity given in Equation \ref{mseequ}, which can be considered as an average approximation error. This is also referred to as the root mean square error in approximating ${(x_i,y_i)}$ by a function $f(x_i, a_1, a_2, \dots, a_n)$.

\begin{equation}
    E=\sqrt{ \frac{1}{n}\sum^n_{i=1} \left[ f(x_i,a_1, a_2, \dots, a_n) - y_i \right]^2}
    \label{mseequ}
\end{equation}

Let's assume that $f(x)$ is in a known class of functions, $C$.  It can be shown that a function $\hat{f}^*$ which is most likely to equal to $f$ will also minimize Equation \ref{mseequhat} among all functions $\hat{f} (x)$ in $C$. This is called the least squares approximation to the data ${(x_i,y_i)}$.

\begin{equation}
    E=\sqrt{ \frac{1}{n}\sum^n_{i=1} \left[ \hat{f}(x_i,a_1, a_2, \dots, a_n) - y_i \right]^2}
    \label{mseequhat}
\end{equation}

Minimizing E is equivalent to minimizing $R^2$, although the minimum values will be different. Thus we seek to minimize Equation \ref{vlsf}, which result in the condition given in Equation \ref{minimize} for $i=1,\dots, n$.

\begin{equation}
    \frac{\partial (R^2)}{\partial a_i}=0
    \label{minimize}
\end{equation}

Let's consider $f(x)=mx+b$ for a linear fit. Thus  we attempt to minimize Equation \ref{linfit}, where $b$ and $m$ are allowed to vary arbitrarily. 
\begin{equation}
R^2=\sum_{i=1}^n\left[mx_i+b-y_i\right]^2
\label{linfit}
\end{equation}

Now, according to Equation  \ref{minimize}, the choices of $b$ and $m$ that minimize $R^2$ satisfy, Equation \ref{part1} and Equation \ref{part2}.
\begin{align}
    \frac{\partial R^2}{\partial b}=0 \label{part1}\\
     \frac{\partial R^2}{\partial m}=0 \label{part2}
\end{align}

Equation \ref{part1} and Equation \ref{part2} result in Equation \ref{solpart1} and Equation \ref{solpart2}.

\begin{equation}
    \frac{\partial R^2}{\partial b}=\sum_{i=1}^n 2 \left[ mx_i+b-y_i \right]
    \label{solpart1}
\end{equation}

\begin{equation}
    \frac{\partial R^2}{\partial m}=\sum_{i=1}^n 2 \left[ mx_i^2+bx_i-x_iy_i \right]
    \label{solpart2}
\end{equation}

From Equation \ref{solpart1} and Equation \ref{solpart2}, we can generate the linear system shown in Equation \ref{linsystem1} which can be represented by the matrix form shown in Equation \ref{matrixform1}. Now, we can solve Equation \ref{matrixform2}, to find values of $a$ and $b$ to obtain the corresponding linear fit of $f(x)=mx+b$.

\begin{align}
\begin{split}
    nb + \left(\sum_{i=1}^n x_i \right) m = \sum_{i=1}^n y_i \\
    \left(\sum_{i=1}^n x_i \right) b + \left(\sum_{i=1}^n x_i^2 \right) m = \sum_{i=1}^n x_i y_i
\end{split}    
\label{linsystem1}
\end{align}

\begin{equation}
    \begin{bmatrix}
    n &  \left(\sum_{i=1}^n x_i \right) \\
   \left(\sum_{i=1}^n x_i \right) & \left(\sum_{i=1}^n x_i^2 \right) 
   \end{bmatrix}
   \begin{bmatrix}
    b \\
    m 
   \end{bmatrix}
   =
   \begin{bmatrix}
    \sum_{i=1}^n y_i \\
    \sum_{i=1}^n x_i y_i 
   \end{bmatrix}
    \label{matrixform1}
\end{equation}

So, 

\begin{equation}
\begin{bmatrix}
    b \\
    m 
   \end{bmatrix}
   =
    \begin{bmatrix}
    n &  \left(\sum_{i=1}^n x_i \right) \\
   \left(\sum_{i=1}^n x_i \right) & \left(\sum_{i=1}^n x_i^2 \right) 
   \end{bmatrix}^{-1}
   \begin{bmatrix}
    \sum_{i=1}^n y_i \\
    \sum_{i=1}^n x_i y_i 
   \end{bmatrix}
    \label{matrixform2}
\end{equation}

\section{Privacy-Preserving Polynomial Model Generation}

Consider a dataset \{$(x_i,y_i) | 1\leq i \leq n\}$, and let
\begin{equation}
 \hat{f}(x)=a_1\varphi_1(x)+a_2\varphi_2(x)+\dots+a_m\varphi_m(x)   
 \label{approxfunc}
\end{equation}
where, $a_1,a_2,\dots,a_m$ are coefficients and $\varphi_1(x), \varphi_2(x),\dots, \varphi_m(x)$ are Chebeshev polynomials of first kind,

\begin{gather}
    \varphi_1(x)=T_0(x)=1 \\
    \varphi_2(x)=T_1(x)=x \\
    \varphi_n(x)=T_{n+1}(x)=2xT_n(x)-T_{n-1}(x)
\label{onesesti}
\end{gather}

Assume that the data $\{x_i\}$ are chosen from an interval $[\alpha, \beta]$. The Chebyshev polynomials can be modified as given in Equation \ref{conditioned1},

\begin{equation}
    \varphi_k(x)=T_{k-1}\left(\frac{2x-\alpha-\beta}{\beta-\alpha}\right)
    \label{conditioned1}
\end{equation}
The approximated function $\hat{f}$ of degree $(m-1)$ can be given by Equation \ref{approxfunc}, where the degree of $(\varphi_k)$ is $k-1$. We will assume the interval $[\alpha, \beta]=[0,1]$ and construct the model accordingly.
According to Equation \ref{conditioned} when $[\alpha, \beta]=[0,1]$, we get Equation \ref{conditioned}.

\begin{equation}
    \varphi_k(x)=T_{k-1}\left(\frac{2x-\alpha-\beta}{\beta-\alpha}\right)=T_{k-1}(2x-1)
    \label{conditioned}
\end{equation}

From Equation \ref{approxfunc} and Equation \ref{conditioned} we have the following equations for $m=4$.

\begin{gather}
    \varphi_1(x)=T_0(2x-1)=1 \\
    \varphi_2(x)=T_1(2x-1)=2x-1 \\
    \varphi_3(x)=T_2(2x-1)=8x^2- 8x + 1 \\
    \varphi_4(x)=T_3(2x-1)=32x^3-48x^2+18x-1
    \label{modifiedapproxy}
\end{gather}

Equation \ref{newapfunc} defines $\hat{f}(x)$ when $m=4$.
\begin{gather}
 \hat{f}(x)=a_1\varphi_1(x)+a_2\varphi_2(x)+a_3\varphi_3(x)+a_4\varphi_4(x)\\
  \hat{f}(x)=a_1(1)+a_2(2x-1)+a_3(8x^2- 8x + 1)+a_4(32x^3-48x^2+18x-1)
 \label{newapfunc}
\end{gather}

Let the actual input be $y_i,$ where $i=1$ to $n$. The error of the approximated input can be determined by Equation \ref{approxierror}.

\begin{equation}
    e_i=\hat{f}(x_i)-y_i
    \label{approxierror}
\end{equation}

We need to determine the values of $a_1, a_2, a_3,$ and $a_4$ in such a way that the errors ($e_i$) are small. In order to determine the best values for $a_1, a_2, a_3,$ and $a_4$, we use the root mean square error given in Equation \ref{rmse}. 

\begin{equation}
    E=\sqrt{\frac{1}{n}\sum_{i=1}^n \left[\hat{f}(x_i)-y_i \right]^2}
    \label{rmse}
\end{equation}

Let's take the least squares fitting of $\hat{f}(x)$ of the class of functions $C$ which minimizes $E$ as $\hat{f}^*(x)$. We can obtain $\hat{f}^*(x)$ by minimizing $E$. Thus we seek to minimize $M(a_1,a_2,a_3,a_4)$ which is given in Equation \ref{minrmse}.

\begin{equation}
    M(a_1,a_2,a_3,a_4)=\sum_{i=1}^n \left[a_1+a_2(2x-1)+a_3(8x^2- 8x + 1)+a_4(32x^3-48x^2+18x-1)-y_i \right]^2
    \label{minrmse}
\end{equation}

The values of $a_1, a_2, a_3,$ and $a_4$ that minimize $M(a_1,a_2,a_3,a_4)$ will satisfy the expressions given in Equation \ref{partial1}, Equation \ref{partial2}, Equation \ref{partial3}, and Equation \ref{partial4}.

\begin{equation}
\resizebox{0.9\hsize}{!}{$
    \frac{\partial M(a_1,a_2,a_3,a_4)}{\partial a_1} = \frac{\partial \left(\sum_{i=1}^n \left[a_1+a_2(2x-1)+a_3(8x^2- 8x + 1)+a_4(32x^3-48x^2+18x-1)-y_i \right]^2\right)}{\partial a_1}= 0 
    $}
    \label{partial1}
\end{equation}

\begin{equation}
\resizebox{0.9\hsize}{!}{$
    \frac{\partial M(a_1,a_2,a_3,a_4)}{\partial a_2} = \frac{\partial \left(\sum_{i=1}^n \left[a_1+a_2(2x-1)+a_3(8x^2- 8x + 1)+a_4(32x^3-48x^2+18x-1)-y_i \right]^2\right)}{\partial a_2}= 0 
    $}
    \label{partial2}
\end{equation}

\begin{equation}
\resizebox{0.9\hsize}{!}{$
    \frac{\partial M(a_1,a_2,a_3,a_4)}{\partial a_3} = \frac{\partial \left(\sum_{i=1}^n \left[a_1+a_2(2x-1)+a_3(8x^2- 8x + 1)+a_4(32x^3-48x^2+18x-1)-y_i \right]^2\right)}{\partial a_3}= 0  
    $}
    \label{partial3}
\end{equation}

\begin{equation}
\resizebox{0.9\hsize}{!}{$
    \frac{\partial M(a_1,a_2,a_3,a_4)}{\partial a_4} = \frac{\partial \left(\sum_{i=1}^n \left[a_1+a_2(2x-1)+a_3(8x^2- 8x + 1)+a_4(32x^3-48x^2+18x-1)-y_i \right]^2\right)}{\partial a_4}= 0 
    $}
    \label{partial4}
\end{equation}

\subsection{Utilizing differential privacy to introduce randomization to the approximation process}

To decide the amount of noise, we have to determine the sensitivity of the noise addition process. Given that we add the noise to the approximated values of $\hat{f(x)}$, the sensitivity $(\Delta f)$ can be defined using Equation \ref{sensitivity}, which is the maximum difference between the highest and the lowest possible output values of $\hat{f(x)}$. Since the input dataset is normalized within the bounds of $0$ and $1$, the minimum possible input or output is 0 while the maximum possible input or output is 1. Therefore, we define the sensitivity of the noise addition process to be $1$.

\begin{gather}
    \Delta f=\lVert max(y_i)-min(y_{i+1}) \rVert_1= (1-0)=1\
    \label{sensitivity}
\end{gather}

Now we add random Laplacian to  each expression given in Equations \ref{partial1} - \ref{partial4}, according to Equation \ref{diffeq2} with a sensitivity ($\Delta f$) of 1 and a privacy budget of $\epsilon$ as shown in Equations \ref{rapartial1},  \ref{rapartial2},  \ref{rapartia3} and  \ref{rapartia4}. In the process of adding the Laplacian noise, we choose Laplacian noise with a position (location) of $0$ as the idea is to keep the local minima of RMSE around $0$ during the process of interpolation.  Here we try to randomize the process of obtaining the local minima of the mean squared error to generate the value for the coefficients $(a_1, a_2, a_3, a_4)$ with randomization. The differentially private Laplacian mechanism can be represented by Equation \ref{diffeq2}, where $\Delta f$ is the sensitivity of the process, and $\epsilon$ is the privacy budget.


\begin{equation}
    PF(D)= \frac{\epsilon}{2\Delta f}e^{-\frac{|x-F(D)|}{\Delta F}}
\label{diffeq2}
\end{equation}

Equation \ref{rapartial1} shows the process of using the Laplacian mechanism to introduce noise to the RMSE minimization of the polynomial interpolation process. Here, we try to introduce an error to the partial derivative of Equation \ref{minrmse} with respect to $a_1$. By doing so,  it guarantees that Equation \ref{rapartial1} contributes with an error to the process of finding the coefficients for $a_1, a_2, a_3,$ and $a_4$, which is given in Equation \ref{vectorA}. Since the sensitivity ($\Delta f$) of the noise addition process is $1$, as defined in Equation \ref{sensitivity}, the scale (spread) of the Laplacian noise is $1/\epsilon$.  We restrict the position ($\mu$) of the Laplacian noise at $0$ as the goal is to achieve the global minima keeping the RMSE at 0 after the randomization.

\begin{equation}
\resizebox{0.9\hsize}{!}{$
    \frac{\partial M(a_1,a_2,a_3,a_4)}{\partial a_1} = \frac{\partial \left(\sum_{i=1}^n \left[a_1+a_2(2x-1)+a_3(8x^2- 8x + 1)+a_4(32x^3-48x^2+18x-1)+ Lap_i(\frac{\Delta f}{\epsilon})-y_i \right]^2\right)}{\partial a_1}= 0
    $}
    \label{rapartial1}    
\end{equation}

After applying the partial derivation on Equation \ref{rapartial1} with respect to $a_1$, we can obtain Equation \ref{rapartial12} which leads to obtaining Equation \ref{rapartial13}.

\begin{equation}
\resizebox{0.9\hsize}{!}{$
    \sum_{i=1}^n 2 \left[a_1+a_2(2x-1)+a_3(8x^2- 8x + 1)+a_4(32x^3-48x^2+18x-1)+ Lap_i(\frac{\Delta f}{\epsilon})-y_i \right]= 0
    $}
    \label{rapartial12}
\end{equation}

Let's use $m_{ij}$ to denote the coefficients, and $b_i$ to represent the constants in the right hand side of the equal symbol in the factorised Equations \ref{rapartial13}, \ref{rapartial23}, \ref{rapartial33}, and \ref{rapartial43}. 

\begin{multline}
     a_1 \underbrace{n}_{m_{11}} +a_2\underbrace{\left(2\left(\sum_{i=1}^n x_i\right)-n\right)}_{m_{12}}+a_3\underbrace{\left(8\left(\sum_{i=1}^n x_i^2\right)-8\left(\sum_{i=1}^n x_i\right)+n\right)}_{m_{13}}+
     \\a_4\underbrace{\left(32\left(\sum_{i=1}^n x_i^3\right)-48\left(\sum_{i=1}^n x_i^2\right)+18\left(\sum_{i=1}^n x_i\right) -n\right)}_{m_{14}}
     =\underbrace{\sum_{i=1}^n y_i -  \sum_{i=1}^n Lap_i\left(\frac{\Delta f}{\epsilon}\right)}_{b_1}
    \label{rapartial13}
\end{multline} 

We repeat the same process of applying noise in Equation \ref{rapartial1} to apply noise to Equation \ref{partial2}.

\begin{equation}
\resizebox{0.9\hsize}{!}{$
    \frac{\partial M(a_1,a_2,a_3,a_4)}{\partial a_2} = \frac{\partial \left(\sum_{i=1}^n \left[a_1+a_2(2x-1)+a_3(8x^2- 8x + 1)+a_4(32x^3-48x^2+18x-1)+Lap_i(\frac{\Delta f}{\epsilon})-y_i \right]^2\right)}{\partial a_2}= 0  
    $}
    \label{rapartial2}
\end{equation}

After solving Equation \ref{rapartial2}, we can obtain Equation \ref{rapartial21} which leads to obtaining Equation \ref{rapartial23}.

\begin{equation}
\resizebox{0.9\hsize}{!}{$
    \sum_{i=1}^n 2  \left[a_1+a_2(2x-1)+a_3(8x^2- 8x + 1)+a_4(32x^3-48x^2+18x-1)+Lap_i(\frac{\Delta f}{\epsilon})-y_i \right](2x_i-1)= 0 
    $}
    \label{rapartial21}
\end{equation}

\begin{multline}
a_1 \underbrace{\left(2\left(\sum_{i=1}^n x_i \right) - n\right)}_{m_{21}} + a_2\underbrace{\left( 4 \left(\sum_{i=1}^n x_i^2 \right) - 4 \left(\sum_{i=1}^n x_i\right) + n  \right)}_{m_{22}}+\\
a_3 \underbrace{\left( 16\left(\sum_{i=1}^n x_i^3 \right) - 24 \left(\sum_{i=1}^n x_i^2 \right)+ 10 \left(\sum_{i=1}^n x_i \right) - n\right)}_{m_{23}}+\\
a_4\underbrace{\left( 64 \left(\sum_{i=1}^n x_i^4 \right) -128 \left(\sum_{i=1}^n x_i^3 \right) + 84 \left(\sum_{i=1}^n x_i^2 \right)- 20 \left( \sum_{i=1}^n x_i\right)+ n\right)}_{m_{24}}
\\
= \underbrace{2\left(\sum_{i=1}^n x_i y_i\right)-\sum_{i=1}^n y_i - \left(  2\left(\sum_{i=1}^n x_i Lap_i\left(\frac{\Delta f}{\epsilon}\right)\right)-\sum_{i=1}^n Lap_i\left(\frac{\Delta f}{\epsilon}\right)  \right)  }_{b_2}
    \label{rapartial23}
\end{multline}

Similarly, we can obtain Equation \ref{rapartial33}  and Equation \ref{rapartial43} after introducing the calibrated Laplacian noise to Equation \ref{partial3} and Equation \ref{partial4} respectively.

\begin{equation}
\resizebox{0.9\hsize}{!}{$
    \frac{\partial M(a_1,a_2,a_3,a_4)}{\partial a_3} = \frac{\partial \left(\sum_{i=1}^n \left[a_1+a_2(2x-1)+a_3(8x^2- 8x + 1)+a_4(32x^3-48x^2+18x-1)+Lap_i(\frac{\Delta f}{\epsilon})-y_i \right]^2\right)}{\partial a_3}= 0
    $}
    \label{rapartia3}
\end{equation}

\begin{equation}
\resizebox{0.9\hsize}{!}{$
    \sum_{i=1}^n 2 \left[a_1+a_2(2x-1)+a_3(8x^2- 8x + 1)+a_4(32x^3-48x^2+18x-1)+Lap_i(\frac{\Delta f}{\epsilon})-y_i \right](8x^2- 8x + 1)= 0
    $}
    \label{rapartia32}
\end{equation}

\begin{multline}
   a_1\underbrace{\left(8\left(\sum_{i=1}^n x_i^2 \right) -8 \left(\sum_{i=1}^n x_i \right)+n\right)}_{m_{31}} + a_2\underbrace{\left( 16 \left(\sum_{i=1}^n x_i^3 \right)- 24 \left(\sum_{i=1}^n x_i^2 \right) + 10 \left(\sum_{i=1}^n x_i \right) - n \right)}_{m_{32}}+\\
   a_3\underbrace{\left(64 \left(\sum_{i=1}^n x_i^4 \right) - 128 \left(\sum_{i=1}^n x_i^3 \right)+80 \left(\sum_{i=1}^n x_i^2 \right) -16 \left(\sum_{i=1}^n x_i \right) + n \right)}_{m_{33}}+\\
   a_4 \underbrace{\left(256 \left( \sum_{i=1}^n x_i^5 \right) - 640 \left( \sum_{i=1}^n x_i^4\right) + 560\left(\sum_{i=1}^n x_i^3\right) - 200 \left(\sum_{i=1}^n x_i^2 \right) + 26 \left( \sum_{i=1}^n x_i\right) - n\right)}_{m_{34}}
   \\ = \underbrace{\begin{aligned} 8\left(\sum_{i=1}^n x_i^2 y_i \right) - 8\left(\sum_{i=1}^n x_i Lap_i\left(\frac{\Delta f}{\epsilon}\right) \right) + \sum_{i=1}^n y_i - \\
   \left(8\left(\sum_{i=1}^n x_i^2 Lap_i\left(\frac{\Delta f}{\epsilon}\right) \right) - 8\left(\sum_{i=1}^n x_i Lap_i\left(\frac{\Delta f}{\epsilon}\right) \right) + \sum_{i=1}^n Lap_i\left(\frac{\Delta f}{\epsilon}\right) \right)\end{aligned}}_{b_3} 
    \label{rapartial33}
\end{multline}

\begin{equation}
\resizebox{0.9\hsize}{!}{$
    \frac{\partial M(a_1,a_2,a_3,a_4)}{\partial a_4} = \frac{\partial \left(\sum_{i=1}^n \left[a_1+a_2(2x-1)+a_3(8x^2- 8x + 1)+a_4(32x^3-48x^2+18x-1)+Lap_i(\frac{\Delta f}{\epsilon})-y_i \right]^2\right)}{\partial a_4}=0
    \label{rapartia4}
    $}
\end{equation}

\begin{equation}
\resizebox{0.9\hsize}{!}{$
    \sum_{i=1}^n 2 \left[a_1+a_2(2x-1)+a_3(8x^2- 8x + 1)+a_4(32x^3-48x^2+18x-1)+Lap_i(\frac{\Delta f}{\epsilon})-y_i \right](32x^3-48x^2+18x-1)= 0
    $}
    \label{rapartial42}
\end{equation}
 
\begin{multline}
      a_1 \underbrace{\left( 32\left( \sum_{i=1}^n x_i^3 \right) - 48\left(\sum_{i=1}^n x_i^2 \right) + 18\left(\sum_{i=1}^n x_i \right) - n\right)}_{m_{41}}+\\
      a_2\underbrace{\left( 64 \left(\sum_{i=1}^n x_i^4 \right) - 128 \left(\sum_{i=1}^n x_i^3 \right)+ 84\left(\sum_{i=1}^n x_i^2\right) - 20 \left(\sum_{i=1}^n x_i \right) + n \right)}_{m_{42}}+\\
      a_3\underbrace{\left( 256 \left(\sum_{i=1}^n x_i^5 \right) - 640 \left(\sum_{i=1}^n x_i^4 \right) + 560 \left(\sum_{i=1}^n x_i^3 \right) - 200 \left(\sum_{i=1}^n x_i^2\right) + 26\left(\sum_{i=1}^n x_i \right) - n \right)}_{m_{43}} +\\
      a_4 \underbrace{\left(1024 \left(\sum_{i=1}^n x_i^6 \right) - 3072 \left( \sum_{i=1}^n x_i^5 \right) + 3456 \left( \sum_{i=1}^n x_i^4 \right) - 1792 \left(\sum_{i=1}^n x_i^3 \right) + 420 \left( \sum_{i=1}^n x_i^2\right) - 36 \left(\sum_{i=1}^n x \right) + n\right)}_{m_{44}}\\
      =\underbrace{\begin{aligned}32 \left(\sum_{i=1}^n x_i^3 y_i \right)- 48 \left(\sum_{i=1}^n x_i^2 y_i \right) + 18 \left(\sum_{i=1}^n x_i y_i \right) - \sum_{i=1}^n y_i - \\
      \left(32 \left(\sum_{i=1}^n x_i^3 Lap_i\left(\frac{\Delta f}{\epsilon}\right) \right)- 48 \left(\sum_{i=1}^n x_i^2 Lap_i\left(\frac{\Delta f}{\epsilon}\right) \right) + 18 \left(\sum_{i=1}^n x_i Lap_i\left(\frac{\Delta f}{\epsilon}\right) \right) - \sum_{i=1}^n Lap_i\left(\frac{\Delta f}{\epsilon}\right)\right)
      \end{aligned}}_{b_4}
    \label{rapartial43}
\end{multline}

Let's consider the coefficients ($m_{ij}$) and the constants ($b_i$) in the Equations \ref{rapartial13}, \ref{rapartial23}, \ref{rapartial33}, and \ref{rapartial43}. Using $m_{ij}$ and $b_i$ we can form a linear system which can be denoted by Equation \ref{linsys}. 

\begin{equation}
\resizebox{0.08\hsize}{!}{$
    CA=B
    $}
    \label{linsys}
\end{equation}

Where,  

\begin{equation}
\resizebox{0.35\hsize}{!}{$
    C=
    \begin{bmatrix}
    m_{11} &  m_{12} & m_{13} & m_{14} \\
    m_{21} & m_{22} & m_{23} & m_{24} \\
    m_{31} & m_{32} & m_{33} & m_{34} \\
    m_{41} & m_{42} & m_{43} & m_{44}
    
   \end{bmatrix}
   $}
    \label{matM}
\end{equation}

\begin{equation}
A=\left[a_1, a_2, a_3, a_4\right]^T
    \label{vectorA}
\end{equation}

\begin{equation}
B=\left[b_1, b_2, b_3, b_4\right]^T
\label{vectorB}
\end{equation}


\begin{thebibliography}{10}
\expandafter\ifx\csname url\endcsname\relax
  \def\url#1{\texttt{#1}}\fi
\expandafter\ifx\csname urlprefix\endcsname\relax\def\urlprefix{URL }\fi
\expandafter\ifx\csname href\endcsname\relax
  \def\href#1#2{#2} \def\path#1{#1}\fi

\bibitem{de2016iot}
G.~De~Francisci~Morales, A.~Bifet, L.~Khan, J.~Gama, W.~Fan, Iot big data
  stream mining, in: Proceedings of the 22nd ACM SIGKDD International
  Conference on Knowledge Discovery and Data Mining, ACM, 2016, pp. 2119--2120.

\bibitem{bertino2008survey}
E.~Bertino, D.~Lin, W.~Jiang, A survey of quantification of privacy preserving
  data mining algorithms, in: Privacy-preserving data mining, Springer, 2008,
  pp. 183--205.

\bibitem{zhang2016privacy}
Q.~Zhang, L.~T. Yang, Z.~Chen, Privacy preserving deep computation model on
  cloud for big data feature learning, IEEE Transactions on Computers 65~(5)
  (2016) 1351--1362.

\bibitem{wen2018scheduling}
Y.~Wen, J.~Liu, W.~Dou, X.~Xu, B.~Cao, J.~Chen,
  \href{http://www.sciencedirect.com/science/article/pii/S0167739X17307379}{Scheduling
  workflows with privacy protection constraints for big data applications on
  cloud}, Future Generation Computer Systems\href
  {http://dx.doi.org/10.1016/j.future.2018.03.028}
  {\path{doi:10.1016/j.future.2018.03.028}}.
\newline\urlprefix\url{http://www.sciencedirect.com/science/article/pii/S0167739X17307379}

\bibitem{xue2011distributed}
M.~Xue, P.~Papadimitriou, C.~Ra{\"\i}ssi, P.~Kalnis, H.~K. Pung, Distributed
  privacy preserving data collection, in: International Conference on Database
  Systems for Advanced Applications, Springer, 2011, pp. 93--107.

\bibitem{backes2016profile}
M.~Backes, P.~Berrang, O.~Goga, K.~P. Gummadi, P.~Manoharan, On profile
  linkability despite anonymity in social media systems, in: Proceedings of the
  2016 ACM on Workshop on Privacy in the Electronic Society, ACM, 2016, pp.
  25--35.

\bibitem{yang2017efficient}
K.~Yang, Q.~Han, H.~Li, K.~Zheng, Z.~Su, X.~Shen, An efficient and fine-grained
  big data access control scheme with privacy-preserving policy, IEEE Internet
  of Things Journal 4~(2) (2017) 563--571.

\bibitem{vatsalan2017privacy}
D.~Vatsalan, Z.~Sehili, P.~Christen, E.~Rahm, Privacy-preserving record linkage
  for big data: Current approaches and research challenges, in: Handbook of Big
  Data Technologies, Springer, 2017, pp. 851--895.

\bibitem{chen2005random}
K.~Chen, L.~Liu, \href{https://corescholar.libraries.wright.edu/knoesis/916/}{A
  random rotation perturbation approach to privacy preserving data
  classification}.
\newline\urlprefix\url{https://corescholar.libraries.wright.edu/knoesis/916/}

\bibitem{chen2011geometric}
K.~Chen, L.~Liu, Geometric data perturbation for privacy preserving outsourced
  data mining, Knowledge and Information Systems 29~(3) (2011) 657--695.

\bibitem{li2015towards}
J.~Li, D.~Lin, A.~C. Squicciarini, J.~Li, C.~Jia, Towards privacy-preserving
  storage and retrieval in multiple clouds, IEEE Transactions on Cloud
  Computing 5~(3) (2017) 499--509.
\newblock \href {http://dx.doi.org/10.1109/TCC.2015.2485214}
  {\path{doi:10.1109/TCC.2015.2485214}}.

\bibitem{kerschbaum2017searchable}
F.~Kerschbaum, M.~H{\"a}rterich, Searchable encryption to reduce encryption
  degradation in adjustably encrypted databases, in: IFIP Annual Conference on
  Data and Applications Security and Privacy, Springer, 2017, pp. 325--336.

\bibitem{gai2016privacy}
K.~Gai, M.~Qiu, H.~Zhao, J.~Xiong, Privacy-aware adaptive data encryption
  strategy of big data in cloud computing, in: Cyber Security and Cloud
  Computing (CSCloud), 2016 IEEE 3rd International Conference on, IEEE, 2016,
  pp. 273--278.

\bibitem{agrawal2005framework}
S.~Agrawal, J.~R. Haritsa, A framework for high-accuracy privacy-preserving
  mining, in: Data Engineering, 2005. ICDE 2005. Proceedings. 21st
  International Conference on, IEEE, 2005, pp. 193--204.

\bibitem{xu2014building}
H.~Xu, S.~Guo, K.~Chen, Building confidential and efficient query services in
  the cloud with rasp data perturbation, IEEE transactions on knowledge and
  data engineering 26~(2) (2014) 322--335.

\bibitem{muralidhar1999general}
K.~Muralidhar, R.~Parsa, R.~Sarathy, A general additive data perturbation
  method for database security, management science 45~(10) (1999) 1399--1415.

\bibitem{fox2015randomized}
J.~A. Fox, Randomized response and related methods: Surveying Sensitive Data,
  Vol.~58, SAGE Publications, 2015.

\bibitem{machanavajjhala2015designing}
A.~Machanavajjhala, D.~Kifer, Designing statistical privacy for your data,
  Communications of the ACM 58~(3) (2015) 58--67.

\bibitem{niu2014achieving}
B.~Niu, Q.~Li, X.~Zhu, G.~Cao, H.~Li, Achieving k-anonymity in privacy-aware
  location-based services, in: INFOCOM, 2014 Proceedings IEEE, IEEE, 2014, pp.
  754--762.

\bibitem{zhang2016designing}
Y.~Zhang, W.~Tong, S.~Zhong, On designing satisfaction-ratio-aware truthful
  incentive mechanisms for $ k $-anonymity location privacy, IEEE Transactions
  on Information Forensics and Security 11~(11) (2016) 2528--2541.

\bibitem{machanavajjhala2006diversity}
A.~Machanavajjhala, J.~Gehrke, D.~Kifer, M.~Venkitasubramaniam, l-diversity:
  Privacy beyond k-anonymity, in: Data Engineering, 2006. ICDE'06. Proceedings
  of the 22nd International Conference on, IEEE, 2006, pp. 24--24.

\bibitem{zhang2007information}
L.~Zhang, S.~Jajodia, A.~Brodsky, Information disclosure under realistic
  assumptions: Privacy versus optimality, in: Proceedings of the 14th ACM
  conference on Computer and communications security, ACM, 2007, pp. 573--583.

\bibitem{ganta2008composition}
S.~R. Ganta, S.~P. Kasiviswanathan, A.~Smith, Composition attacks and auxiliary
  information in data privacy, in: Proceedings of the 14th ACM SIGKDD
  international conference on Knowledge discovery and data mining, ACM, 2008,
  pp. 265--273.

\bibitem{wong2011can}
R.~C.-W. Wong, A.~W.-C. Fu, K.~Wang, P.~S. Yu, J.~Pei, Can the utility of
  anonymized data be used for privacy breaches?, ACM Transactions on Knowledge
  Discovery from Data (TKDD) 5~(3) (2011) 16.

\bibitem{dwork2009differential}
C.~Dwork, The differential privacy frontier, in: Theory of Cryptography
  Conference, Springer, 2009, pp. 496--502.

\bibitem{mohammed2011differentially}
N.~Mohammed, R.~Chen, B.~Fung, P.~S. Yu, Differentially private data release
  for data mining, in: Proceedings of the 17th ACM SIGKDD international
  conference on Knowledge discovery and data mining, ACM, 2011, pp. 493--501.

\bibitem{friedman2010data}
A.~Friedman, A.~Schuster, Data mining with differential privacy, in:
  Proceedings of the 16th ACM SIGKDD international conference on Knowledge
  discovery and data mining, ACM, 2010, pp. 493--502.

\bibitem{wang2015outsourcing}
W.~Wang, L.~Chen, Q.~Zhang, Outsourcing high-dimensional healthcare data to
  cloud with personalized privacy preservation, Computer Networks 88 (2015)
  136--148.

\bibitem{qin2016heavy}
Z.~Qin, Y.~Yang, T.~Yu, I.~Khalil, X.~Xiao, K.~Ren, Heavy hitter estimation
  over set-valued data with local differential privacy, in: Proceedings of the
  2016 ACM SIGSAC Conference on Computer and Communications Security, ACM,
  2016, pp. 192--203.

\bibitem{kellaris2014differentially}
G.~Kellaris, S.~Papadopoulos, X.~Xiao, D.~Papadias, Differentially private
  event sequences over infinite streams, Proceedings of the VLDB Endowment
  7~(12) (2014) 1155--1166.

\bibitem{aggarwal2005k}
C.~C. Aggarwal, On k-anonymity and the curse of dimensionality, in: Proceedings
  of the 31st international conference on Very large data bases, VLDB
  Endowment, 2005, pp. 901--909.

\bibitem{mivule2013comparative}
K.~Mivule, C.~Turner, A comparative analysis of data privacy and utility
  parameter adjustment, using machine learning classification as a gauge,
  Procedia Computer Science 20 (2013) 414--419.

\bibitem{kieseberg2018security}
P.~Kieseberg, E.~Weippl, Security challenges in cyber-physical production
  systems, in: International Conference on Software Quality, Springer, 2018,
  pp. 3--16.

\bibitem{balandina2015iot}
E.~Balandina, S.~Balandin, Y.~Koucheryavy, D.~Mouromtsev, Iot use cases in
  healthcare and tourism, in: Business Informatics (CBI), 2015 IEEE 17th
  Conference on, Vol.~2, IEEE, 2015, pp. 37--44.

\bibitem{sridhar2012cyber}
S.~Sridhar, A.~Hahn, M.~Govindarasu, et~al., Cyber-physical system security for
  the electric power grid., Proceedings of the IEEE 100~(1) (2012) 210--224.

\bibitem{fernando2016consumer}
R.~Fernando, R.~Ranchal, B.~An, L.~B. Othman, B.~Bhargava, Consumer oriented
  privacy preserving access control for electronic health records in the cloud,
  in: Cloud Computing (CLOUD), 2016 IEEE 9th International Conference on, IEEE,
  2016, pp. 608--615.

\bibitem{liu2012cyber}
J.~Liu, Y.~Xiao, S.~Li, W.~Liang, C.~P. Chen, Cyber security and privacy issues
  in smart grids, IEEE Communications Surveys \& Tutorials 14~(4) (2012)
  981--997.

\bibitem{bertino2016data}
E.~Bertino, Data privacy for iot systems: concepts, approaches, and research
  directions, in: Big Data (Big Data), 2016 IEEE International Conference on,
  IEEE, 2016, pp. 3645--3647.

\bibitem{wang2014performance}
X.~Wang, J.~Zhang, E.~M. Schooler, M.~Ion, Performance evaluation of
  attribute-based encryption: Toward data privacy in the iot, in:
  Communications (ICC), 2014 IEEE International Conference on, IEEE, 2014, pp.
  725--730.

\bibitem{kirkham2015privacy}
T.~Kirkham, A.~Sinha, N.~Parlavantzas, B.~Kryza, P.~Fremantle, K.~Kritikos,
  B.~Aziz, Privacy aware on-demand resource provisioning for iot data
  processing, in: International Internet of Things Summit, Springer, 2015, pp.
  87--95.

\bibitem{aldeen2015comprehensive}
Y.~A. A.~S. Aldeen, M.~Salleh, M.~A. Razzaque, A comprehensive review on
  privacy preserving data mining, SpringerPlus 4~(1) (2015) 694.

\bibitem{verykios2004state}
V.~S. Verykios, E.~Bertino, I.~N. Fovino, L.~P. Provenza, Y.~Saygin,
  Y.~Theodoridis, State-of-the-art in privacy preserving data mining, ACM
  Sigmod Record 33~(1) (2004) 50--57.

\bibitem{razaque2017privacy}
A.~Razaque, S.~S. Rizvi, Privacy preserving model: a new scheme for auditing
  cloud stakeholders, Journal of Cloud Computing 6~(1) (2017) 7.

\bibitem{wu2016scalable}
D.~Wu, B.~Yang, R.~Wang, Scalable privacy-preserving big data aggregation
  mechanism, Digital Communications and Networks 2~(3) (2016) 122--129.

\bibitem{razaque2016triangular}
A.~Razaque, S.~S. Rizvi, Triangular data privacy-preserving model for
  authenticating all key stakeholders in a cloud environment, Computers \&
  Security 62 (2016) 328--347.

\bibitem{agrawal2000privacy}
R.~Agrawal, R.~Srikant, Privacy-preserving data mining, in: ACM Sigmod Record,
  Vol.~29, ACM, 2000, pp. 439--450.

\bibitem{datta2004random}
S.~Datta, On random additive perturbation for privacy preserving data mining,
  Ph.D. thesis, University of Maryland, Baltimore County (2004).

\bibitem{liu2006random}
K.~Liu, H.~Kargupta, J.~Ryan, Random projection-based multiplicative data
  perturbation for privacy preserving distributed data mining, IEEE
  Transactions on knowledge and Data Engineering 18~(1) (2006) 92--106.

\bibitem{zhong2012mu}
J.~Zhong, V.~Mirchandani, P.~Bertok, J.~Harland, $\mu$-fractal based data
  perturbation algorithm for privacy protection., in: PACIS, 2012, p. 148.

\bibitem{du2003using}
W.~Du, Z.~Zhan, Using randomized response techniques for privacy-preserving
  data mining, in: Proceedings of the ninth ACM SIGKDD international conference
  on Knowledge discovery and data mining, ACM, 2003, pp. 505--510.

\bibitem{estivill1999data}
V.~Estivill-Castro, L.~Brankovic, Data swapping: Balancing privacy against
  precision in mining for logic rules, in: DaWaK, Vol.~99, Springer, 1999, pp.
  389--398.

\bibitem{aggarwal2004condensation}
C.~C. Aggarwal, P.~S. Yu, A condensation approach to privacy preserving data
  mining, in: EDBT, Vol.~4, Springer, 2004, pp. 183--199.

\bibitem{domingo2002practical}
J.~Domingo-Ferrer, J.~M. Mateo-Sanz, Practical data-oriented microaggregation
  for statistical disclosure control, IEEE Transactions on Knowledge and data
  Engineering 14~(1) (2002) 189--201.

\bibitem{wong2006alpha}
R.~C.-W. Wong, J.~Li, A.~W.-C. Fu, K.~Wang, ($\alpha$, k)-anonymity: an
  enhanced k-anonymity model for privacy preserving data publishing, in:
  Proceedings of the 12th ACM SIGKDD international conference on Knowledge
  discovery and data mining, ACM, 2006, pp. 754--759.

\bibitem{li2007t}
N.~Li, T.~Li, S.~Venkatasubramanian, t-closeness: Privacy beyond k-anonymity
  and l-diversity, in: Data Engineering, 2007. ICDE 2007. IEEE 23rd
  International Conference on, IEEE, 2007, pp. 106--115.

\bibitem{cao2011castle}
J.~Cao, B.~Carminati, E.~Ferrari, K.-L. Tan, Castle: Continuously anonymizing
  data streams, IEEE Transactions on Dependable and Secure Computing 8~(3)
  (2011) 337--352.

\bibitem{dwork2014algorithmic}
C.~Dwork, A.~Roth, et~al., The algorithmic foundations of differential privacy,
  Foundations and Trends{\textregistered} in Theoretical Computer Science
  9~(3--4) (2014) 211--407.

\bibitem{dwork2008differential}
C.~Dwork, Differential privacy: A survey of results, in: International
  Conference on Theory and Applications of Models of Computation, Springer,
  2008, pp. 1--19.

\bibitem{kairouz2014extremal}
P.~Kairouz, S.~Oh, P.~Viswanath, Extremal mechanisms for local differential
  privacy, in: Advances in neural information processing systems, 2014, pp.
  2879--2887.

\bibitem{erlingsson2014rappor}
{\'U}.~Erlingsson, V.~Pihur, A.~Korolova, Rappor: Randomized aggregatable
  privacy-preserving ordinal response, in: Proceedings of the 2014 ACM SIGSAC
  conference on computer and communications security, ACM, 2014, pp.
  1054--1067.

\bibitem{cormode2018privacy}
G.~Cormode, S.~Jha, T.~Kulkarni, N.~Li, D.~Srivastava, T.~Wang, Privacy at
  scale: Local differential privacy in practice, in: Proceedings of the 2018
  International Conference on Management of Data, ACM, 2018, pp. 1655--1658.

\bibitem{wold1987principal}
S.~Wold, K.~Esbensen, P.~Geladi, Principal component analysis, Chemometrics and
  intelligent laboratory systems 2~(1-3) (1987) 37--52.

\bibitem{scholz2006maximum}
F.~W. Scholz,
  \href{https://onlinelibrary.wiley.com/doi/abs/10.1002/0471667196.ess1571.pub2}{Maximum
  Likelihood Estimation}, Wiley Online Library, 2006.
\newblock \href {http://dx.doi.org/10.1002/0471667196.ess1571.pub2}
  {\path{doi:10.1002/0471667196.ess1571.pub2}}.
\newline\urlprefix\url{https://onlinelibrary.wiley.com/doi/abs/10.1002/0471667196.ess1571.pub2}

\bibitem{aggarwal2008general}
C.~C. Aggarwal, S.~Y. Philip, A general survey of privacy-preserving data
  mining models and algorithms, in: Privacy-preserving data mining, Springer,
  2008, pp. 11--52.

\bibitem{chen2005privacy}
K.~Chen, L.~Liu, Privacy preserving data classification with rotation
  perturbation, in: Data Mining, Fifth IEEE International Conference on, IEEE,
  2005, pp. 1--4.

\bibitem{huang2005deriving}
Z.~Huang, W.~Du, B.~Chen, Deriving private information from randomized data,
  in: Proceedings of the 2005 ACM SIGMOD international conference on Management
  of data, ACM, 2005, pp. 37--48.

\bibitem{domingo2017steered}
J.~Domingo-Ferrer, J.~Soria-Comas, Steered microaggregation: A unified
  primitive for anonymization of data sets and data streams, in: Data Mining
  Workshops (ICDMW), 2017 IEEE International Conference on, IEEE, 2017, pp.
  995--1002.

\bibitem{zhang2015proximity}
X.~Zhang, W.~Dou, J.~Pei, S.~Nepal, C.~Yang, C.~Liu, J.~Chen, Proximity-aware
  local-recoding anonymization with mapreduce for scalable big data privacy
  preservation in cloud, IEEE transactions on computers 64~(8) (2015)
  2293--2307.

\bibitem{aggarwal2008static}
C.~C. Aggarwal, P.~S. Yu, On static and dynamic methods for condensation-based
  privacy-preserving data mining, ACM Transactions on Database Systems (TODS)
  33~(1) (2008) 2.

\bibitem{chamikaraprocal}
M.~A.~P. Chamikara, P.~Bertok, D.~Liu, S.~Camtepe, I.~Khalil, Efficient data
  perturbation for privacy preserving and accurate data stream mining,
  Pervasive and Mobile Computing\href
  {http://dx.doi.org/10.1016/j.pmcj.2018.05.003}
  {\path{doi:10.1016/j.pmcj.2018.05.003}}.

\bibitem{chamikara2019efficient}
M.~A.~P. Chamikara, P.~Bertok, D.~Liu, S.~Camtepe, I.~Khalil, Efficient privacy
  preservation of big data for accurate data mining, Information Sciences\href
  {http://dx.doi.org/10.1016/j.ins.2019.05.053}
  {\path{doi:10.1016/j.ins.2019.05.053}}.

\bibitem{xu2008privacy}
Y.~Xu, K.~Wang, A.~W.-C. Fu, R.~She, J.~Pei, Privacy-preserving data stream
  classification, Privacy-Preserving Data Mining (2008) 487--510.

\bibitem{li2007hiding}
F.~Li, J.~Sun, S.~Papadimitriou, G.~A. Mihaila, I.~Stanoi, Hiding in the crowd:
  Privacy preservation on evolving streams through correlation tracking, in:
  Data Engineering, 2007. ICDE 2007. IEEE 23rd International Conference on,
  IEEE, 2007, pp. 686--695.

\bibitem{mason2002chebyshev}
J.~C. Mason, D.~C. Handscomb, Chebyshev polynomials, Chapman and Hall/CRC,
  2002.

\bibitem{weisstein2002least}
E.~W. Weisstein, Least squares fitting.

\bibitem{chan2012differentially}
T.-H.~H. Chan, M.~Li, E.~Shi, W.~Xu, Differentially private continual
  monitoring of heavy hitters from distributed streams, in: International
  Symposium on Privacy Enhancing Technologies Symposium, Springer, 2012, pp.
  140--159.

\bibitem{wang2016using}
Y.~Wang, X.~Wu, D.~Hu, Using randomized response for differential privacy
  preserving data collection., in: EDBT/ICDT Workshops, Vol. 1558, 2016.

\bibitem{hammad2003scheduling}
M.~A. Hammad, M.~J. Franklin, W.~G. Aref, A.~K. Elmagarmid, Scheduling for
  shared window joins over data streams, in: Proceedings of the 29th
  international conference on Very large data bases-Volume 29, VLDB Endowment,
  2003, pp. 297--308.

\bibitem{abadi2016deep}
M.~Abadi, A.~Chu, I.~Goodfellow, H.~B. McMahan, I.~Mironov, K.~Talwar,
  L.~Zhang, Deep learning with differential privacy, in: Proceedings of the
  2016 ACM SIGSAC Conference on Computer and Communications Security, ACM,
  2016, pp. 308--318.

\bibitem{baheti2011cyber}
R.~Baheti, H.~Gill, Cyber-physical systems, The impact of control technology
  12~(1) (2011) 161--166.

\bibitem{alhayajneh2018biometric}
A.~Alhayajneh, A.~Baccarini, G.~Weiss, T.~Hayajneh, A.~Farajidavar, Biometric
  authentication and verification for medical cyber physical systems,
  Electronics 7~(12) (2018) 436.

\bibitem{witten2016data}
I.~H. Witten, E.~Frank, M.~A. Hall, C.~J. Pal, Data Mining: Practical machine
  learning tools and techniques, Morgan Kaufmann, 2016.

\bibitem{okkalioglu2015survey}
B.~D. Okkalioglu, M.~Okkalioglu, M.~Koc, H.~Polat, A survey: deriving private
  information from perturbed data, Artificial Intelligence Review 44~(4) (2015)
  547--569.

\bibitem{lessmann2015benchmarking}
S.~Lessmann, B.~Baesens, H.-V. Seow, L.~C. Thomas, Benchmarking
  state-of-the-art classification algorithms for credit scoring: An update of
  research, European Journal of Operational Research 247~(1) (2015) 124--136.

\bibitem{scholkopf1999advances}
B.~Sch{\"o}lkopf, C.~J. Burges, A.~J. Smola, Advances in kernel methods:
  support vector learning, MIT press, 1999.

\bibitem{quinlan1993c4}
J.~R. Quinlan, C4. 5: Programming for machine learning, Morgan Kauffmann 38.

\bibitem{howell2016fundamental}
D.~C. Howell, Fundamental statistics for the behavioral sciences, Nelson
  Education, 2016.

\bibitem{gavert2005fastica}
H.~G{\"a}vert, J.~Hurri, J.~S{\"a}rel{\"a}, A.~Hyv{\"a}rinen,
  \href{https://research.ics.aalto.fi/ica/fastica/}{The fastica package for
  matlab}, Lab Comput Inf Sci Helsinki Univ. Technol.
\newline\urlprefix\url{https://research.ics.aalto.fi/ica/fastica/}

\bibitem{zarzoso2006fast}
V.~Zarzoso, P.~Comon, M.~Kallel, How fast is fastica?, in: 2006 14th European
  Signal Processing Conference, IEEE, 2006, pp. 1--5.

\bibitem{shukla2016benchmarking}
A.~Shukla, Y.~Simmhan, Benchmarking distributed stream processing platforms for
  iot applications, in: Technology Conference on Performance Evaluation and
  Benchmarking, Springer, 2016, pp. 90--106.

\end{thebibliography}
\end{document}